\documentclass[runningheads]{latex/llncs}
\usepackage[T1]{fontenc}
\usepackage{circledsteps}
\usepackage{hyperref}

\usepackage{caption}
\usepackage{subcaption}
\usepackage{multirow}
\usepackage{graphicx}
\usepackage{tikz}
\usetikzlibrary{
    positioning,
    shapes,
    arrows,
    fit,
}

\usepackage[printonlyused,nolist]{acronym}
\usepackage[capitalise,nameinlink]{cleveref}
\usepackage{siunitx}
\usepackage{orcidlink}

\makeatletter
\renewcommand*\AC@acs[1]{%
    \expandafter\AC@get\csname fn@#1\endcsname\@firstoftwo{#1}}
\makeatother

\begin{acronym}[AAPCS64]
    \acro{aapcs64}[AAPCS64]{Procedure Call Standard for the \acl{aarch64}}
    \acro{aarch64}[AArch64]{\acs{arm} 64-bit Architecture}
    \acro{ai}[AI]{Artificial Intelligence}
    \acro{aiba}[AIBA]{An Automated Intra-cycle Behavioral Analysis for SystemC-based design exploration}
    \acro{algol60}[ALGOL~60]{Algorithmic Language 1960}
    \acro{amd}[AMD]{Advanced Micro Devices}
    \acro{aoa}[AoA]{ARM-on-ARM}
    \acro{api}[API]{Application Programming Interface}
    \acro{arm}[ARM]{Advanced \acs{risc} Machines}
    \acro{asil}[ASIL]{Automotive Safety Integrity Level}
    \acro{ast}[AST]{Abstract Syntax Tree}
    \acro{avp64}[AVP64]{\acs{arm}v8 \acl{vp}}
    \acro{bar}[BAR]{Base Address Register}
    \acro{bb}[BB]{Basic Block}
    \acro{bl}[BL]{Branch-With-Link}
    \acro{bp}[BP]{Breakpoint}
    \acro{can}[CAN]{Controller Area Network}
    \acro{ci}[CI]{Continous Integration}
    \acro{cicd}[CI/CD]{Continuous Integration/Continuous Delivery}
    \acro{clint}[CLINT]{Core-Local Interrupt Controller}
    \acro{cpu}[CPU]{Central Processing Unit}
    \acro{crc}[CRC]{Cyclic Redundancy Check}
    \acro{csv}[CSV]{Character-Separated Values}
    \acro{db}[DB]{Database}
    \acro{dbt}[DBT]{Dynamic Binary Translation}
    \acro{dbms}[DBMS]{\Acl{db} Management System}
    \acro{ddr}[DDR]{Double Data Rate}
    \acro{des}[DES]{Discrete Event Simulation}
    \acro{dla}[DLA]{Deep Learning Accelerator}
    \acro{dma}[DMA]{Direct Memory Access}
    \acro{dmi}[DMI]{Direct Memory Interface}
    \acro{ds}[DS]{Developer Studio}
    \acro{dwarf}[DWARF]{Debugging With Arbitrary Record Formats}
    \acro{ecu}[ECU]{Electronic Control Unit}
    \acro{eda}[EDA]{Electronic Design Automation}
    \acro{eembc}[EEMBC]{Embedded Microprocessor Benchmark Consortium}
    \acro{etrace}[etrace]{Execution Trace}
    \acro{el}[EL]{Exception Level}
    \acro{elf}[ELF]{Executable and Linkable Format}
    \acro{elog}[elog]{Execution Log}
    \acro{esa}[ESA]{European Space Agency}
    \acro{esl}[ESL]{Electronic System Level}
    \acro{fd}[fd]{file descriptor}
    \acro{fig}[FIG]{Fault Injection in glibc}
    \acro{fmi}[FMI]{Functional Mock-up Interface}
    \acro{fmi-ls-bus}[FMI-LS-BUS]{FMI Layered Standard for Network Communication}
    \acro{fmu}[FMU]{Functional Mock-up Unit}
    \acro{fp}[FP]{Frame Pointer}
    \acro{fpga}[FPGA]{Field Programmable Gate Array}
    \acro{fss}[FSS]{Full-System Simulator}
    \acro{fvp}[FVP]{Fixed Virtual Platform}
    \acro{gcc}[GCC]{GNU Compiler Collection}
    \acro{gdb}[GDB]{GNU Debugger}
    \acro{gic}[GIC]{Generic Interrupt Controller}
    \acro{got}[GOT]{Global Offset Table}
    \acro{gpio}[GPIO]{General Purpose Input/Output}
    \acro{gpu}[GPU]{Graphics Processing Unit}
    \acro{gui}[GUI]{Graphical User Interface}
    \acro{hart}[hart]{Hardware Thread}
    \acro{html}[HTML]{HyperText Markup Language}
    \acro{hw}[HW]{Hardware}
    \acro{ibm}[IBM]{International Business Machines}
    \acro{id}[ID]{identifier}
    \acro{ieee}[IEEE]{Institute of Electrical and Electronics Engineers}
    \acro{io}[I/O]{Input/Output}
    \acro{iommu}[IOMMU]{Input-Output Memory Management Unit}
    \acro{iot}[IoT]{Intenet of Things}
    \acro{iova}[IOVA]{IO Virtual Address}
    \acroplural{iova}[IOVA]{IO Virtual Address}
    \acrodefplural{iova}[IOVAs]{IO Virtual Addresses}
    \acro{ip}[IP]{Internet Protocol}
    \acro{ipa}[IPA]{Intermediate Physical Address}
    \acrodefplural{ipa}[IPAs]{Intermediate Physical Addresses}
    \acro{ir}[IR]{Intermediate Representation}
    \acro{irq}[IRQ]{Interrupt Request}
    \acro{isa}[ISA]{Instruction-Set Architecture}
    \acro{isa-bus}[ISA]{Industry Standard Architecture}
    \acro{iss}[ISS]{Instruction-Set Simulator}
    \acro{jit}[JIT]{Just-In-Time}
    \acro{kvm}[KVM]{Kernel Virtual Machine}
    \acro{l4vecu}[L4 vECU]{Level~4 Virtual Electronic Control Unit}
    \acro{lr}[LR]{Link Register}
    \acro{lt}[LT]{Loosely-Timed}
    \acro{mcdc}[MC/DC]{Modified Condition/Decision Coverage}
    \acro{mcu}[MCU]{Microcontroller Unit}
    \acro{mips}[MIPS]{Million Instructions Per Second}
    \acro{miso}[MISO]{Master Input Slave Output}
    \acro{mmio}[MMIO]{Memory-Mapped Input/Output}
    \acro{mmu}[MMU]{Memory Management Unit}
    \acro{mnist}[MNIST]{Modified National Institute of Standards and Technology}
    \acro{mosi}[MOSI]{Master Output Slave Input}
    \acro{msi}[MSI]{Message Signalled Interrupt}
    \acro{nas}[NAS]{\acs{nasa} Advanced Supercomputing}
    \acro{nasa}[NASA]{National Space Agency}
    \acro{nistt}[NISTT]{A Non-Intrusive SystemC-TLM 2.0 Tracing Tool}
    \acro{nn}[NN]{Neural Network}
    \acro{nop}[NOP]{No Operation}
    \acro{npb}[NPB]{\acs{nas} Parallel Benchmarks}
    \acro{nvdla}[NVDLA]{NVIDIA \acl*{dla}}
    \acro{openmp}[OpenMP]{Open Multi-Processing}
    \acro{os}[OS]{Operating System}
    \acro{pa}[PA]{Physical Address}
    \acrodefplural{pa}[PAs]{Physical Addresses}
    \acro{pc}[PC]{Program Counter}
    \acro{pccts}[PCCTS]{Purdue Compiler Construction Tool Set}
    \acro{pci}[PCI]{Peripheral Component Interconnect}
    \acro{pcie}[PCIe]{PCI Express}
    \acro{pid}[PID]{process ID}
    \acro{plic}[PLIC]{Platform-Level Interrupt Controller}
    \acro{plt}[PLT]{Procedure Linkage Table}
    \acro{pmu}[PMU]{Performance Monitoring Unit}
    \acro{pthread}[p\-thread]{\acs{posix} Thread}
    \acro{posix}[POSIX]{Portable Operating System Interface}
    \acro{qemu}[QEMU]{Quick Emulator}
    \acro{ram}[RAM]{Random-Access Memory}
    \acro{risc}[RISC]{Reduced Instruction Set Computer}
    \acro{rsp}[RSP]{Remote Serial Protocol}
    \acro{rtc}[RTC]{Real-Time Clock}
    \acro{rtf}[RTF]{Real-Time Factor}
    \acro{rtl}[RTL]{Register-Transfer Level}
    \acro{rtos}[RTOS]{real-time operating system}
    \acro{sata}[SATA]{Serial AT Attachment}
    \acro{sd}[SD]{Secure Digital}
    \acro{sdhci}[SDHCI]{\acs{sd} Host Controller Interface}
    \acro{simd}[SIMD]{Single Instruction, Multiple Data}
    \acro{smmu}[SMMU]{System Memory Management Unit}
    \acro{soc}[SoC]{System-on-a-Chip}
    \acrodefplural{soc}[SoCs]{Systems-on-Chips}
    \acro{sp}[SP]{Stack Pointer}
    \acro{spi}[SPI]{Serial Peripheral Interface}
    \acro{sql}[SQL]{Structured Query Language}
    \acro{ssd}[SSD]{Solid State Drive}
    \acro{sw}[SW]{Software}
    \acro{tb}[TB]{Translation Block}
    \acro{tcg}[TCG]{Tiny Code Generator}
    \acro{tcp}[TCP]{Transmission Control Protocol}
    \acro{td}[TD]{Temporal Decoupling}
    \acro{tfl}[TFLite]{TensorFlow Lite}
    \acro{tlm}[TLM]{Transaction-Level Modeling}
    \acro{tpu}[TPU]{Tensor Processing Unit}
    \acro{uart}[UART]{Universal Asynchronous Receiver/Transmitter}
    \acro{unix}[UNIX]{Uniplexed Information and Computing Service}
    \acro{va}[VA]{Virtual Address}
    \acrodefplural{va}[VAs]{Virtual Addresses}
    \acro{vcd}[VCD]{Value Change Dump}
    \acro{vcml}[VCML]{Virtual Components Modeling Library}
    \acro{vcpu}[vCPU]{virtual CPU}
    \acro{vfio}[VFIO]{Virtual Function I/O}
    \acro{vhe}[VHE]{Virtualization Host Extensions}
    \acro{vpci}[vPCI(e)]{virtual PCI(e)}
    \acro{viper}[VIPER]{Virtual Platform Explorer}
    \acro{vsp}[VSP]{VCML Session Protocol}
    \acro{vpci}[vPCI]{virtual PCI}
    \acro{vm}[VM]{Virtual Machine}
    \acro{vp}[VP]{Virtual Platform}
    \acro{vt-d}[VT-d]{Virtualization Technology for Directed I/O}
    \acro{wfi}[WFI]{Wait For Interrupt}
\end{acronym}

\definecolor{rwth}   {RGB}{  0  84 159}
\definecolor{rwth-75}{RGB}{ 64 127 183}
\definecolor{rwth-50}{RGB}{142 186 229}
\definecolor{rwth-25}{RGB}{199 221 242}
\definecolor{rwth-10}{RGB}{232 241 250}

\definecolor{black}   {RGB}{  0   0   0}
\definecolor{black-75}{RGB}{100 101 103}
\definecolor{black-50}{RGB}{156 158 159}
\definecolor{black-25}{RGB}{207 209 210}
\definecolor{black-10}{RGB}{236 237 237}

\definecolor{magenta}   {RGB}{227   0 102}
\definecolor{magenta-75}{RGB}{233  96 136}
\definecolor{magenta-50}{RGB}{241 158 177}
\definecolor{magenta-25}{RGB}{249 210 218}
\definecolor{magenta-10}{RGB}{253 238 240}

\definecolor{yellow}   {RGB}{255 237   0}
\definecolor{yellow-75}{RGB}{255 240  85}
\definecolor{yellow-50}{RGB}{255 245 155}
\definecolor{yellow-25}{RGB}{255 250 209}
\definecolor{yellow-10}{RGB}{255 253 238}

\definecolor{petrol}   {RGB}{  0  97 101}
\definecolor{petrol-75}{RGB}{ 45 127 131}
\definecolor{petrol-50}{RGB}{125 164 167}
\definecolor{petrol-25}{RGB}{191 208 209}
\definecolor{petrol-10}{RGB}{230 236 236}

\definecolor{turkis}   {RGB}{  0 152 161}
\definecolor{turkis-75}{RGB}{  0 177 183}
\definecolor{turkis-50}{RGB}{137 204 207}
\definecolor{turkis-25}{RGB}{202 231 231}
\definecolor{turkis-10}{RGB}{235 246 246}

\definecolor{grun}   {RGB}{ 87 171  39}
\definecolor{grun-75}{RGB}{141 192  96}
\definecolor{grun-50}{RGB}{184 214 152}
\definecolor{grun-25}{RGB}{221 235 206}
\definecolor{grun-10}{RGB}{242 247 236}

\definecolor{maigrun}   {RGB}{189 205   0}
\definecolor{maigrun-75}{RGB}{208 217  92}
\definecolor{maigrun-50}{RGB}{224 230 154}
\definecolor{maigrun-25}{RGB}{240 243 208}
\definecolor{maigrun-10}{RGB}{249 250 237}

\definecolor{orange}   {RGB}{246 168   0}
\definecolor{orange-75}{RGB}{250 190  80}
\definecolor{orange-50}{RGB}{253 212 143}
\definecolor{orange-25}{RGB}{254 234 201}
\definecolor{orange-10}{RGB}{255 247 234}

\definecolor{rot}   {RGB}{204   7  30}
\definecolor{rot-75}{RGB}{216  92  65}
\definecolor{rot-50}{RGB}{230 150 121}
\definecolor{rot-25}{RGB}{243 205 187}
\definecolor{rot-10}{RGB}{250 235 227}

\definecolor{bordeaux}   {RGB}{161  16  53}
\definecolor{bordeaux-75}{RGB}{182  82  86}
\definecolor{bordeaux-50}{RGB}{205 139 135}
\definecolor{bordeaux-25}{RGB}{229 197 192}
\definecolor{bordeaux-10}{RGB}{245 232 229}

\definecolor{lila}   {RGB}{122 111 172}
\definecolor{lila-75}{RGB}{155 145 193}
\definecolor{lila-50}{RGB}{188 181 215}
\definecolor{lila-25}{RGB}{222 218 235}
\definecolor{lila-10}{RGB}{242 240 247}

\definecolor{violett}   {RGB}{ 97  33  88}
\definecolor{violett-75}{RGB}{131  78 117}
\definecolor{violett-50}{RGB}{168 133 158}
\definecolor{violett-25}{RGB}{210 192 205}
\definecolor{violett-10}{RGB}{237 229 234}

\tikzset{/csteps/fill color=white, /csteps/outer color=bordeaux-75, /csteps/inner color=black}

\usepackage{pgfplots}
\pgfplotsset{compat=1.18}
\usepackage{pgfplotstable}

\usetikzlibrary{patterns}
\pgfplotsset{
    y axis style/.style={
        yticklabel style=#1,
        ylabel style=#1,
        y axis line style=#1,
        ytick style=#1
    },
    General/.style={
        font=\footnotesize,
        width=\linewidth,
        height=4.2cm,
        xtick pos=left,
        xtick align=outside,
        ytick pos=left,
        ytick align=outside,
        ymajorgrids=true,
        grid style=dashed,
        legend cell align={left},
        style=thin,
    },
    BarConfig/.style={
        General,
        ymin=0,
        enlarge x limits=0.1,
        xtick=data,
        xticklabel style={rotate=45, anchor=east},
        xlabel shift=-6pt,
        bar width=8pt,
    },
    ScatterConfig/.style={
      General,
      only marks,
      xtick=data,
      xticklabel style={rotate=90, anchor=east},
    }
}
\newcommand{\legendposition}{north west}
\Crefname{figure}{Figure}{Figures}
\usepackage{color}

\usepackage[pages=some]{background}
\begin{document}
\title{Bridging the Gap: Physical PCI Device Integration Into SystemC-TLM Virtual Platforms}
\titlerunning{Physical PCI Device Integration Into SystemC-TLM Virtual Platforms}
%
\author{
    Nils~Bosbach\inst{1}\orcidlink{0000-0002-2284-949X} \and
    Rebecca~Pelke\inst{1}\orcidlink{0000-0001-5156-7072} \and
    Niko~Zurstraßen\inst{1}\orcidlink{0000-0003-3434-2271} \and
    Jan~Henrik~Weinstock\inst{2}\orcidlink{0009-0008-0902-7652} \and
    Lukas~Jünger\inst{2}\orcidlink{0000-0001-9149-1690} \and
    Rainer~Leupers\inst{1}\orcidlink{0000-0002-6735-3033}
}
\authorrunning{N. Bosbach et al.}
%
\institute{
    RWTH Aachen University, Germany \and
    MachineWare GmbH, Germany
}

\newcommand\copyrightnotice{%
    \backgroundsetup{opacity=1, scale=1, angle=0, contents={
            \color{black}%
            \begin{tikzpicture}[remember picture,overlay]%
                \node[anchor=north,yshift=-10pt,text=gray] at (current page.north) {\shortstack[c]{\large PREPRINT - accepted by the \textit{25th International Conference on Embedded Computer Systems:}\\\textit{Architectures, Modeling and Simulation (SAMOS XXV)}}};
            \end{tikzpicture}%
        }%
    }%
    \BgThispage%
}

\maketitle
\copyrightnotice

%
%
\acused{cpu}
\acused{tlm}
\begin{abstract}
    In today's technology-driven world, early-stage software development and testing are crucial.
    \acp{vp} have become indispensable tools for this purpose as they serve as a platform to execute and debug the unmodified target software at an early design stage.
    With the increasing complexity of software, especially in areas like \ac{ai} applications, \acp{vp} need to provide high simulation speed to ensure the target software executes within a reasonable time.
    Hybrid simulation, which combines virtual models with real hardware, can improve the performance of \acp{vp}.

    This paper introduces a novel approach for integrating real \ac{pci} devices into SystemC-\ac{tlm}-2.0-based \acp{vp}.
    The embedded \ac{pci} devices enable high performance, easy integration, and allow introspection for analysis and optimization.

    To illustrate the practical application of our approach, we present a case study where we integrate Google Coral's Edge \ac{tpu} into an ARM-based \ac{vp}.
    The integration allows efficient execution of \ac{ai} workloads, accelerating simulation speeds by up to \SI{480}{\texttimes} while eliminating the need for complex virtual device models.
    Beyond accelerating \ac{ai}-workload execution, our framework enables driver development, regression testing across architectures, and device communication analysis.
    Our findings demonstrate that embedding \ac{pci} devices into SystemC simulations significantly enhances their capabilities, paving the way for more effective virtual prototyping.

\end{abstract}
\acresetall
\acused{qemu}
\acused{cpu}
\section{Introduction}
In an era where the complexity of both systems and \ac{sw} continues to grow exponentially, the need for early-stage software development and testing is paramount.
\acp{vp} are indispensable tools for architecture exploration, \ac{hw}/\ac{sw} co-design, and design verification and testing~\cite{vinco_saga_2012}.
However, their effectiveness heavily depends on their simulation speed.

One factor that contributes to the complexity of \acp{vp} is the incorporation of compute-intensive machine-learning applications in the target \ac{sw}.
Predictions forecast that the global embedded \ac{ai} market size will grow from USD~\num{8.79} billion in 2023 to USD~\num{21.93} billion in 2030, which highlights the relevance of this type of workload~\cite{grand_view_research_embedded_nodate}.
A solution that enables fast execution of \ac{ai} workloads in \acp{vp} is hybrid simulation.
Thereby, virtual models and real \ac{hw} devices are combined.
Since machine learning accelerators are often external devices that communicate with a host machine via interfaces like \ac{pci}, they are well-suited for hybrid simulation.

In this paper, we present a way of integrating real \ac{pci}(e)\footnote{The acronym \acs{pci}(e) is used in the following to refer to both \acs{pci} and \acs{pcie} devices} devices into a SystemC-Transaction-Level-Modeling~(\acs{tlm})\acused{tlm}-based \ac{vp}.
SystemC~\cite{systemc-2023} is the industry standard for system-level simulation.
It is extended by \ac{tlm}, which enables communication between models without simulating protocol details.
Standardized interfaces allow model exchange and reuse even between vendors.

In the past, previous simulators have already demonstrated the advantages of integrating \ac{pci}(e) devices.
A popular example is \ac{qemu}~\cite{bellard_qemu_2005}.
\ac{qemu} is a versatile simulator that is capable of executing \ac{sw} compiled for various architectures on several host architectures.
It has a \ac{pci}(e) pass-through feature that allows to pass \ac{pci}(e) devices to the simulator.
However, the drawback of \ac{qemu} compared to a SystemC-\ac{tlm}-based simulation is the absence of standardized interfaces.
This lack of standardization makes it challenging to adapt simulations for specialized use cases, integrate additional models, or reuse existing ones.
Furthermore, \ac{qemu} is neither timed nor deterministic, which are essential properties for reliable analysis.

\begin{figure}[t]
    \centering
    \input{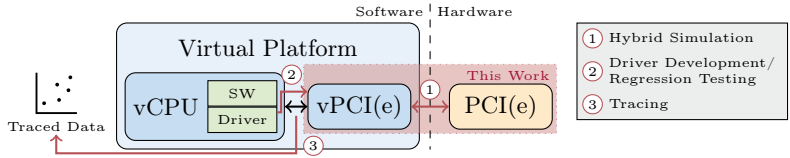}
    \vspace{-1em}
    \caption{Application scenarios of the \acs{pci}(e) integration.}
    \label{fig:overview}
    \vspace{-2em}
\end{figure}

\Cref{fig:overview} shows a general overview of a \ac{vpci} device and its integration into a \ac{vp}.
Our approach can be used in the following scenarios:

\begin{enumerate}
    \item[\Circled{1}] \textbf{Hybrid Simulation}: Our \ac{vp} integration can be used to embed a real device into a virtual system.
        It eliminates the need to develop a virtual model of the \ac{pci}(e) device, which speeds up the development process.
        The target \ac{sw} can access both, virtual models and the \ac{hw} \ac{pci}(e) device.

    \item[\Circled{2}] \textbf{Driver Development}: Device drivers for different \ac{cpu} architectures can be developed on one machine.
        Additionally, a single machine can run multiple \acp{vp} to simulate different architectures.
        This simplifies regression testing since the \ac{pci}(e) card does not need to be connected to different machines.

    \item[\Circled{3}] \textbf{Device-Communication Analysis}: The accesses from the \ac{vcpu} to the \ac{pci}(e) device can be traced and analyzed.
        Furthermore, the interrupt behavior of the \ac{pci}(e) device can be evaluated.
\end{enumerate}

In a case study in \cref{sec:casestudy}, we showcase how our approach can integrate Google Coral's Edge \ac{tpu} into an ARM-based \ac{vp}.

\section{Background}
This chapter explains the background information that is needed to follow our approach.
A short overview of the development of \ac{pci}(e) and its working principles is given in \cref{sec:background-pci}.
In \cref{sec:background-vfio}, Linux's \ac{vfio} driver is presented.
The SystemC-based \ac{vcml} is presented in \cref{sec:background-vcml}, which is our used simulation environment.

\subsection{\acf{pci}}
\label{sec:background-pci}

The \ac{pci} bus specification 1.0 was published by Intel in 1992 as a successor of the \acl{isa-bus} bus.
Over the years, multiple revisions have been published to meet the demands of faster and more complex \ac{hw}.

Nowadays, \ac{pcie} is the standard for extension cards that can be connected to a motherboard, like graphics cards, network cards, or storage devices, such as \acp{ssd}.
Additionally, \ac{pci} is used to connect on-chip devices.

To connect a \ac{pci}(e) card to the system bus, a \ac{pci} host bridge (\textit{\acs{pci} host}) is required.
A card can contain multiple endpoints, so-called \textit{functions}, which can be accessed by individual drivers.
Each function has a \textit{configuration} memory space with a standardized layout, which can be accessed by the \ac{cpu}.

The configuration space contains six \acp{bar} (\ac{bar}0-\ac{bar}5).
Those registers contain information on additional memory regions the device has.
During bus enumeration, the \ac{os} needs to program those registers to map the available memory region to either the memory or \ac{io} space.~\cite{anderson_pci_1999}

Most \ac{pci}(e) devices have a \ac{dma} engine to directly access the host memory without \ac{cpu} interaction.
In a typical procedure, the \ac{cpu} places data in memory that can be directly accessed by the \ac{pci}(e) card.
The \ac{cpu} then configures the card via the mapped memory areas.

To notify the \ac{cpu}, interrupts are available.
\ac{pci} supports four interrupt lines that are shared between all devices~\cite{pci_sig_pci_1998}.
Interrupt sharing can lead to reduced performance because the \ac{os} needs to call all handlers associated with an interrupt every time an interrupt is signaled~\cite{nguyen_4_2008}.
Since \ac{pci}~2.3, \acp{msi} can be used if supported by the device \cite{pci_sig_pci_2003}.
Using \ac{msi}, the \ac{os} programs an address to which the device can write to trigger an interrupt.
In \ac{pci}~3.0, \ac{msi}-X was added which allows more interrupts per device and independent configuration.
For \ac{pcie} devices, the support of either \ac{msi} or \ac{msi}-X is mandatory.

\subsection{\acf{vfio}}
\label{sec:background-vfio}

\ac{vfio} is a Linux user-level driver framework originally developed by Cisco in 2010~\cite{williamson_vfio_2012}.
It has been part of the Linux kernel since version 3.6.0, which was released in 2012~\cite{linus_torvalds_linux-kernel_2012}.
It is a device driver to get raw access to \ac{pci}(e) devices from a user-space process.
Accesses to the configuration space are virtualized.
The \ac{mmio} regions of the \ac{pci}(e) device can directly be mapped into the address space of a process.

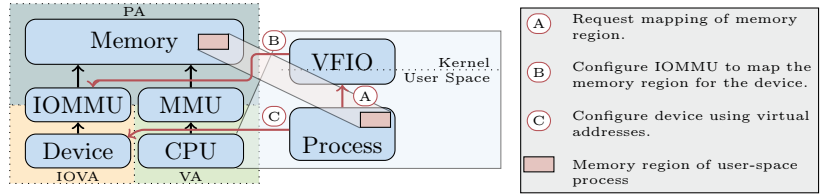
\begin{figure}[t]
    \centering
    \tikzstyle{nodetype} = [rectangle, rounded corners, minimum width=1.0cm, minimum height=.45cm, text centered, text width=1.2cm, draw=black, fill=rwth-25, font={\footnotesize}, inner sep=0.1cm]
\tikzstyle{isock}    = [rectangle, minimum width=.15cm, minimum height=.15cm, draw=black, fill=black, inner sep=0pt]
\tikzstyle{tsock}    = [isock, fill=white]
\tikzstyle{isocki}   = [isock, fill=rwth, draw=rwth]
\tikzstyle{tsocki}   = [isocki, fill=white]
\tikzstyle{isockp}   = [isock, fill=grun, draw=grun]
\tikzstyle{tsockp}   = [isockp, fill=white]
\tikzstyle{lbl}      = [font={\tiny}, inner sep=1pt]
\tikzstyle{socklbl}  = [lbl]

\pgfdeclarelayer{background}
\pgfdeclarelayer{middle}
\pgfdeclarelayer{front}
\pgfsetlayers{background,middle,main,front}

\begin{tikzpicture}[node distance=.5cm and .5cm]
    \begin{scope}[name prefix=memory-]
        \node[nodetype, fill=rwth-25, minimum width=2.9cm, minimum height=0.6cm] (main) {Memory};
        \begin{pgfonlayer}{front}
            \node[nodetype, fill=bordeaux-25, anchor=east, minimum width=0.2cm, text width=0.2cm, minimum height=0.15cm, sharp corners] (mem) at ([xshift=-0.2cm]main.east) {};
        \end{pgfonlayer}
    \end{scope}

    \begin{scope}[name prefix=mmu-]
        \node[nodetype, fill=rwth-25, anchor=north east] at ([yshift=-0.3cm]memory-main.south east) (main) {MMU};
    \end{scope}

    \begin{scope}[name prefix=iommu-]
        \node[nodetype, fill=rwth-25, anchor=north west] at (memory-main.west|-mmu-main.north) (main) {IOMMU};
    \end{scope}

    \begin{scope}[name prefix=cpu-]
        \node[nodetype, fill=rwth-25, anchor=north] at ([yshift=-0.15cm]mmu-main.south) (main) {CPU};
    \end{scope}

    \begin{scope}[name prefix=device-]
        \node[nodetype, fill=rwth-25, anchor=north west] at (iommu-main.west|-cpu-main.north) (main) {Device};
    \end{scope}

    \begin{scope}[name prefix=prog-]
        \node[nodetype, fill=rwth-25, text height=0.5cm, anchor=south west] at ([xshift=0.6cm, yshift=0.1cm]cpu-main.south east) (main) {Process};
        \begin{pgfonlayer}{front}
            \node[nodetype, fill=bordeaux-25, anchor=north east, minimum width=0.2cm, text width=0.2cm, minimum height=0.15cm, sharp corners] (mem) at ([xshift=-0.06cm, yshift=-0.06cm]main.north east) {};
        \end{pgfonlayer}
    \end{scope}

    \begin{scope}[name prefix=vfio-]
        \node[nodetype, fill=rwth-25, minimum height=0.6cm, anchor=south] at ([yshift=+0.3cm]prog-main.north) (main) {VFIO};
        \draw[dotted] ([xshift=-0.1cm]main.base-|main.west) -- ([xshift=1.4cm]main.base-|main.east)
        node [lbl, anchor=south east, xshift=-0.1cm] (lbl-kernel) {Kernel}
        node [lbl, anchor=north east, xshift=-0.1cm] (lbl-user) {User Space};
    \end{scope}

    \draw[fill=bordeaux-10, semitransparent] (prog-mem.south west) -- (memory-mem.south west) -- (memory-mem.north east) -- (prog-mem.north east);

    \node[lbl, anchor=north] at (cpu-main.south) (lbl-virtual) {\acs{va}};
    \node[lbl, anchor=north] at (device-main.south) (lbl-device) {\acs{iova}};
    \node[lbl, anchor=south] at (memory-main.north) (lbl-physical) {\acs{pa}};

    \draw[->, thick] (cpu-main.north) -- (mmu-main.south);
    \draw[->, thick] (device-main.north) -- (iommu-main.south);
    \draw[->, thick] (mmu-main.north) -- (mmu-main.north|-memory-main.south);
    \draw[->, thick] (iommu-main.north) -- (iommu-main.north|-memory-main.south);

    \begin{pgfonlayer}{middle}
        \draw[fill=rwth-10, semitransparent] ([xshift=-0.1cm]cpu-main.north east)
        -- ([xshift=-0.1cm, yshift=+0.1cm]vfio-main.north west)
        -| ([xshift=+1.4cm, yshift=-0.1cm]prog-main.south east)
        -- ([xshift=-0.1cm, yshift=-0.1cm]prog-main.south west)
        -- ([xshift=-0.1cm]cpu-main.south east);
    \end{pgfonlayer}

    \begin{pgfonlayer}{background}
        \draw[dotted, fill=orange-25] (memory-main.south|-mmu-main.center) rectangle ([xshift=-0.2cm, yshift=-0.2cm]device-main.south west) node (leftend) {};
        \draw[dotted, fill=grun-25] (memory-main.south|-mmu-main.center) rectangle ([xshift=+0.2cm, yshift=-0.2cm]cpu-main.south east) node (rightend) {};
        \draw[dotted, fill=petrol-25] (leftend|-iommu-main.center) rectangle ([yshift=+0.2cm]rightend|-memory-main.north);
    \end{pgfonlayer}

    \draw[->, thick, bordeaux-75] (prog-main.north) -- node[lbl, anchor=west, xshift=0.1cm] {\Circled{A}} (vfio-main.south);
    \draw[->, thick, bordeaux-75, rounded corners=2pt] ([yshift=0.1cm]vfio-main.west) -| node[lbl, anchor=south, pos=0.2] {\Circled{B}} ++(-0.5cm, -0.35cm) -| ([xshift=0.2cm]iommu-main.north);
    \draw[->, thick, bordeaux-75, rounded corners=2pt] ([yshift=0.05cm]prog-main.west|-device-main.north) -- node[lbl, anchor=south, pos=0.1] {\Circled{C}} ([xshift=0.1cm, yshift=0.05cm]device-main.north east) -- ([xshift=-0.05cm, yshift=-0.05cm]device-main.north east);

    \node[inner sep=0pt, draw=none] at ([xshift=0.3cm, yshift=-0.1cm]current bounding box.north east) (legend-north-west) {};

    \node[lbl, anchor=north west] at (legend-north-west) (legend1) {\Circled{A}};
    \node[lbl, text width=3.2cm, anchor=north west] at ([xshift=0.2cm]legend1.north east) (legend1-lbl) {Request mapping of memory region.};

    \node[lbl, anchor=north] at ([yshift=-0.3cm]legend1.south) (legend2) {\Circled{B}};
    \node[lbl, text width=3.2cm, anchor=north west] at (legend2.north-|legend1-lbl.west) (legend2-lbl) {Configure IOMMU to map the memory region for the device.};

    \node[lbl, anchor=north] at ([yshift=-0.3cm]legend2.south) (legend3) {\Circled{C}};
    \node[lbl, text width=3.2cm, anchor=north west] at (legend3.north-|legend1-lbl.west) (legend3-lbl) {Configure device using virtual addresses.};

    \node[nodetype, fill=bordeaux-25, anchor=north, minimum width=0.2cm, text width=0.2cm, minimum height=0.15cm, sharp corners] (memreg) at ([yshift=-0.3cm]legend3.south) {};
    \node[lbl, text width=3.2cm, anchor=north west] at (memreg.north-|legend1-lbl.west) (memreg-lbl) {Memory region of user-space process};

    \begin{pgfonlayer}{background}
        \node[draw=black, fill=black-10, inner sep=1pt, fit={(legend1) (legend2) (legend3) (memreg) (legend1-lbl) (legend2-lbl) (legend3-lbl) (memreg-lbl)}] {};
    \end{pgfonlayer}


\end{tikzpicture}
    \caption{\acs{iommu} working principle.}
    \label{fig:iommu}
    \vspace{-2em}
\end{figure}

To reroute \ac{dma} accesses of the device to a memory region of the user-space process, an \ac{iommu} of the host system is needed.
The \ac{iommu} translates addresses on \ac{dma} accesses as visualized in \cref{fig:iommu}.
Similar to a \ac{mmu} that translates \acp{va} into \acp{pa} on \ac{cpu} memory accesses, an \ac{iommu} can translate \acp{iova} into \acp{pa} on \ac{dma} accesses of the \ac{pci}(e) device.
During normal operation of a \ac{pci}(e) device, the \ac{iommu} is usually disabled and the device directly uses physical addresses.
\ac{vfio} requires an activated \ac{iommu}.
When the \ac{vfio} driver is activated for a device, a user-space process can use a \ac{vfio} call to map a memory region from its virtual address space into the \ac{io} virtual address space of the \ac{pci}(e) device~\Circled{A}.
\ac{vfio} translates the \acp{va} of the memory region into \acp{pa} using the process's page tables.
The \ac{iommu} is then configured to perform the same mapping as the \ac{mmu} for the memory region~\Circled{B}.
Once the mapping has been set up, the user-space process can pass \acp{va} to the \ac{pci}(e) device~\Circled{C}.
The device will be able to access the regions using its \ac{dma} because the \ac{iommu} translates the \acp{iova} into the corresponding \acp{pa}.
Different \ac{iommu} implementations exist such as \textit{AMD-Vi}~\cite{amd_inc_amd_2023} from AMD, \textit{\ac{vt-d}}~\cite{intel_intel_2022} from Intel, or ARM's \textit{\ac{smmu}}~\cite{arm_arm_2024}.

\ac{vfio} is the ideal driver to connect a \ac{pci}(e) device to a \ac{vm} or \ac{vp}.
Since it is a device-independent driver that allows raw access, it can be used to expose a \ac{pci}(e) to a simulator or virtual environment and leave the device-specific configuration to the virtual system.

\subsection{\acf{vcml}}
\label{sec:background-vcml}

\ac{vcml} \cite{machineware_machineware-gmbhvcml_2024} is an open-source modeling library that is built on top of SystemC.
It adds commonly used building blocks like specialized \ac{tlm} sockets, \ac{io} peripherals, and registers.
Moreover, the library contains ready-to-use models of devices such as interrupt controllers or Ethernet devices.

A feature that is used in this work is the protocol implementation.
SystemC's \ac{tlm} extensions allow us to model communication between \ac{hw} blocks abstractly \cite{systemc-2023}.
Communication always takes place between an \textit{initiator socket} and a \textit{target socket}.
The sockets are objects that can be placed in modules.
An initiator socket can only communicate with the target socket to which it was \textit{bound} during the simulation setup.
The communication itself is called a \textit{transaction}.
A transaction is always started by the initiator and processed by the corresponding target.
\ac{tlm} supports two different abstraction levels to model communication, \textit{blocking} and \textit{non-blocking} transactions.
They differ in the number of phases that are simulated to send and process a transaction.
In \ac{vcml}, the more abstract \ac{tlm} blocking transport is used because the library targets instruction-accurate modeling where the details of \ac{tlm}'s non-blocking transactions, which are used for cycle-accurate modeling, are not needed.

Based on SystemC's generic \ac{tlm} sockets, \ac{vcml} provides implementations for commonly used communication protocols, such as \ac{uart}, \ac{spi}, \ac{can}, and Ethernet.
The protocols provide \ac{tlm} sockets and \ac{tlm} payload implementations.
Particularly relevant for this work is the \ac{pci} protocol, which can be used to connect a virtual \ac{pci} device to a virtual \ac{pci} host controller.
Based on \ac{tlm}, it provides \ac{pci} initiator and target sockets.
The \ac{pci} device has a \ac{pci} target socket that is connected to a \ac{pci} initiator socket of the \ac{pci} host controller.
The controller can send \ac{pci} transactions to the target to access its memory regions.
To signal an interrupt to the \ac{pci} host, the device uses the backward path of the \ac{pci} target socket.

Multiple commercial \acp{vp} from MachineWare~\cite{bosbach_work--progress_2023,bosbach_high-performance_2025,bosbach_towards_2024}, and open-source \acp{vp}~\cite{junger_armv8_2023} exist that are based on \ac{vcml}.
In the case study in \cref{sec:casestudy}, we use an open-source ARM-based \ac{vp} to showcase our integration.
However, as our model is based on SystemC, it can also be integrated into \acp{vp} that do not use \ac{vcml}.

\section{Related Work}
\label{sec:related-work}

\begin{table}[b]
    \centering
    \vspace{-3em}
    \caption{Comparison between different simulators.}
    \vspace{1em}
    \label{tab:qemu}
    \footnotesize
    \newcolumntype{P}[1]{>{\centering\arraybackslash}p{#1}}
    \newcommand{\tick}{\includegraphics[height=0.25cm]{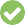}}
    \newcommand{\cross}{\includegraphics[height=0.25cm]{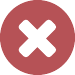}}
    \begin{tabular}{P{2cm}||P{1.3cm}|P{1.5cm}|P{1.6cm}|P{1.5cm}|P{1.7cm}|P{1.5cm}}
        \textbf{}           & \textbf{\ac{vfio}} & \textbf{SystemC} & \textbf{Stand-ardized} & \textbf{Timed} & \textbf{Determi-nistic} & \textbf{Tracing} \\ \hline \hline
        \textbf{\acs{qemu}} & \tick              & \cross           & \cross                 & \cross         & \cross                  & \cross           \\ \hline
        \textbf{This Work}  & \tick              & \tick            & \tick                  & \tick          & \tick                   & \tick
    \end{tabular}
\end{table}

The integration of \ac{pci}(e) devices into other environments like \acp{vm} or simulators is not a new concept.
One of the most well-known simulators supporting \ac{vfio} is \ac{qemu}~\cite{bellard_qemu_2005}, which introduced basic \ac{vfio} support in version 1.3 back in 2012~\cite{bellard_changelog13_2012}.
Since \ac{qemu} was primarily designed for the execution of software compiled for different architectures, it quickly meets its limits when it should be used as a \ac{vp}.
\Cref{tab:qemu} outlines the key differences between our implementation and \ac{qemu}.

Since \ac{qemu} is not based on SystemC, it is not as modular as SystemC-based simulations.
In a SystemC simulation, models are self-contained blocks that communicate via standardized communication points, e.g., \ac{tlm} sockets.
The standardization enables the model exchange, even between different vendors.
Furthermore, tools that rely on the use of SystemC's interfaces can be used~\cite{bosbach_nistt_2022}.

Another drawback of \ac{qemu} compared to SystemC is the lack of a virtual time concept.
In the standard operation mode, \ac{qemu} uses the wall-clock time as reference.
This can lead to non-deterministic behavior.
SystemC, on the other hand, has a concept of a virtual time that is kept track of by the SystemC kernel.
In addition, SystemC simulations are deterministic, meaning that multiple runs of the same simulation with the same inputs will produce exactly the same output.
This is very important for the reproducibility of bugs when \acp{vp} are used for testing.

\section{Approach}
\label{sec:implementation}

In this work, we present a \ac{vpci} device that uses \ac{vfio} to integrate a physical \ac{pci}(e) card that is connected to the host into a \ac{vp}.
The integration is based on \ac{vcml}~\cite{machineware_machineware-gmbhvcml_2024} to benefit from the \ac{tlm}-2.0-based \ac{pci} protocol.

In a case study in \cref{sec:casestudy}, we will show the integration of the \ac{vpci} device into a \ac{vp} in more detail.
For now, we focus on the \ac{vpci} model itself.
Three key features must be supported by the model:

\begin{enumerate}
    \item \textbf{Memory Access}: The \ac{vcpu} model needs to be able to access the physical memory of the \ac{pci}(e) device.

    \item \textbf{\ac{dma}}: \ac{pci}(e) cards can directly access memory using \ac{dma}.
          When such access happens, the virtual memory model needs to be accessed.

    \item \textbf{Interrupt Handling}: \ac{pci} cards can send interrupts.
          When the physical card sends an interrupt, the interrupt needs to be injected into the \ac{vp}.
\end{enumerate}

\subsection{Memory Access}
\label{sec:implementation:memory}

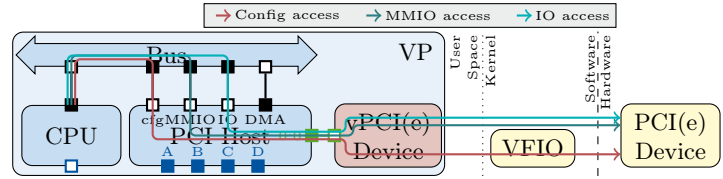
\begin{figure}[!t]
    \centering
    \tikzstyle{nodetype} = [rectangle, rounded corners, minimum width=1.1cm, minimum height=.5cm, text centered, text width=1.1cm, draw=black, fill=rwth-25, font={\footnotesize}]
\tikzstyle{isock}    = [rectangle, thick, minimum width=.15cm, minimum height=.15cm, draw=black, fill=black, inner sep=0pt]
\tikzstyle{tsock}    = [isock, fill=white]
\tikzstyle{isocki}   = [isock, fill=rwth, draw=rwth]
\tikzstyle{tsocki}   = [isocki, fill=white]
\tikzstyle{isockp}   = [isock, fill=grun, draw=grun]
\tikzstyle{tsockp}   = [isockp, fill=white]
\tikzstyle{lbl}      = [font={\tiny}, inner sep=1pt]
\tikzstyle{socklbl}  = [lbl]

\pgfdeclarelayer{background}
\pgfsetlayers{background,main}

\begin{tikzpicture}[node distance=.5cm and .5cm]
    \begin{scope}[name prefix=vfio-]
        \node[nodetype, fill=bordeaux-25, minimum height=0.8cm, anchor=north, text width=1.2cm] (main) {\acs{vpci}(e) Device};
        \node[tsockp] at (main.west) (sock) {};
    \end{scope}

    \begin{scope}[name prefix=pci-host-]
        \node[nodetype, minimum width=2.0cm, text width=2.2cm, minimum height=0.8cm, anchor=east] at ([xshift=-0.3cm]vfio-main.west) (main) {PCI Host};

        \node[tsock, xshift=-0.9cm] at (main.north) (sock_cfg) {};
        \node[socklbl, anchor=north] at (sock_cfg.south) {cfg};

        \node[tsock, xshift=-0.4cm] at (main.north) (sock_mmio) {};
        \node[socklbl, anchor=north] at (sock_mmio.south) {MMIO};

        \node[tsock, xshift=+0.1cm] at (main.north) (sock_io) {};
        \node[socklbl, anchor=north] at (sock_io.south) {IO};

        \node[isock, xshift=+0.6cm] at (main.north) (sock_dma) {};
        \node[socklbl, anchor=north] at (sock_dma.south) {DMA};

        \node[isocki, xshift=-0.7cm] at (main.south) (sock_inta) {};
        \node[socklbl, anchor=base, text=rwth] at ([yshift=0.05cm]sock_inta.north) {A};

        \node[isocki, xshift=-0.3cm] at (main.south) (sock_intb) {};
        \node[socklbl, anchor=base, text=rwth] at ([yshift=0.05cm]sock_intb.north) {B};

        \node[isocki, xshift=+0.1cm] at (main.south) (sock_intc) {};
        \node[socklbl, anchor=base, text=rwth] at ([yshift=0.05cm]sock_intc.north) {C};

        \node[isocki, xshift=+0.5cm] at (main.south) (sock_intd) {};
        \node[socklbl, anchor=base, text=rwth] at ([yshift=0.05cm]sock_intd.north) {D};

        \node[isockp] at (main.east) (sock_pci0_0) {};
        \node[socklbl, anchor=east, text=grun] at (sock_pci0_0.west) {0:0};
    \end{scope}

    \begin{scope}[name prefix=cpu-]
        \node[nodetype, minimum height=0.8cm, anchor=east] at ([xshift=-0.1cm]pci-host-main.west) (main) {CPU};
        \node[isock] at (main.north)(sock) {};
        \node[tsocki] at (main.south)(socki) {};
    \end{scope}

    \node[lbl, fit={(pci-host-main) (cpu-main)}, inner sep=0pt] (peripherals) {};

    \begin{scope}[name prefix=bus-]
        \node[double arrow, draw, minimum height=4.0cm, fill=rwth-25, font={\footnotesize}, anchor=south, inner sep=1pt, double arrow head extend=0.1cm] at ([yshift=+0.5cm]peripherals.north) (main) {Bus};

        \node[isock] at (main.south-|pci-host-sock_cfg) (pci-host-cfg){};
        \node[isock] at (main.south-|pci-host-sock_mmio) (pci-host-mmio){};
        \node[isock] at (main.south-|pci-host-sock_io) (pci-host-io){};
        \node[tsock] at (main.south-|pci-host-sock_dma) (pci-host-dma){};
        \draw[thick] (pci-host-cfg) -- (pci-host-sock_cfg);
        \draw[thick] (pci-host-mmio) -- (pci-host-sock_mmio);
        \draw[thick] (pci-host-io) -- (pci-host-sock_io);
        \draw[thick] (pci-host-dma) -- (pci-host-sock_dma);

        \node[tsock] at (main.south-|cpu-sock) (cpu){};
        \draw[thick] (cpu) -- (cpu-sock);

    \end{scope}

    \draw[draw=grun, thick] (vfio-sock) -- (pci-host-sock_pci0_0);

    \node[anchor=north east] at (vfio-main.east|-bus-main.after tip 2) (lbl_vp) {VP};
    \begin{pgfonlayer}{background}
        \node[nodetype, fill=rwth-10, inner sep=1pt, fit={(vfio-main) (lbl_vp) (pci-host-main) (pci-host-sock_cfg) (bus-main) (bus-main.after tip 2) (cpu-socki)}] (vp) {};
    \end{pgfonlayer}

    \begin{scope}[name prefix=linux-]
        \node[nodetype, fill=yellow-25, text width=0.9cm, anchor=south west] (main) at ([xshift=0.6cm]vfio-main.south-|vp.east) {VFIO};
    \end{scope}south

    \begin{scope}[name prefix=pci-]
        \node[nodetype, fill=yellow-25, anchor=south west] at ([xshift=0.6cm]linux-main.south east) (main) {PCI(e) Device};
    \end{scope}

    \begin{pgfonlayer}{background}
        \draw[dotted] ([xshift=0.5cm]vp.south-|vp.east) -- ([xshift=0.5cm]vp.north-|vp.east)
        node[lbl,rotate=90,anchor=south east] {\shortstack[r]{User\\Space}}
        node[lbl,rotate=90,anchor=north east] {Kernel};
        \draw[dashed] ([xshift=0.3cm]vp.south-|linux-main.east) -- ([xshift=0.3cm]vp.north-|linux-main.east)
        node[lbl,rotate=90,anchor=south east] {Software}
        node[lbl,rotate=90,anchor=north east] {Hardware};
    \end{pgfonlayer}

    \draw[->, bordeaux-75, thick, rounded corners=2pt] ([xshift=0.05cm]cpu-sock.center) |- ([yshift=0.1cm]bus-pci-host-cfg.center) |- ([xshift=0.15cm, yshift=-0.05cm]vfio-main.west) |- ([yshift=-0.25cm]pci-main.west) ;

    \draw[->, petrol-75, thick, rounded corners=2pt] ([xshift=0.00cm]cpu-sock.center) |- ([yshift=0.15cm]bus-pci-host-mmio.center) |- ([xshift=0.25cm, yshift=+0.00cm]vfio-main.west) |- ([yshift=+0.05cm]pci-main.west|-linux-main.north) ;

    \draw[->, turkis-75, thick, rounded corners=2pt] ([xshift=-0.05cm]cpu-sock.center) |- ([yshift=0.15cm]bus-pci-host-io.center) |- ([xshift=0.15cm, yshift=+0.05cm]vfio-main.west) |- ([yshift=+0.15cm]pci-main.west|-linux-main.north) ;

    \node[draw=none, fit={(current bounding box)}, inner sep=0pt] (drawing) {};

    \draw[->, bordeaux-75, thick] ([xshift=-2.0cm, yshift=0.2cm] drawing.north) node (legend-start){} -- ++(0.2cm, 0cm) node[lbl, text=black, anchor=west] {Config access};
    \draw[->, petrol-75, thick] (legend-start-|drawing.north) -- ++(0.2cm, 0cm) node[lbl, text=black, anchor=west] {MMIO access};
    \draw[->, turkis-75, thick] ([xshift=+2.0cm] legend-start-|drawing.north) -- ++(0.2cm, 0cm) node[lbl, text=black, anchor=west] (legend-end) {IO access};
    \begin{pgfonlayer}{background}
        \node[draw=black, fill=black-10, inner sep=1pt, fit={(legend-start) (legend-end)}] {};
    \end{pgfonlayer}

\end{tikzpicture}
    \caption{Memory access implementation.}
    \label{fig:bar}
    \vspace{-2em}
\end{figure}

\ac{pci}(e) devices contain memory that the \ac{cpu} can read from and write to.
As described in \cref{sec:background-pci}, a \ac{pci}(e) device at least contains a configuration memory, that contains device and vendor information.
The memory access from the \ac{cpu} to the \ac{pci}(e) device is depicted in \cref{fig:bar}.
The memory regions of the device are mapped into the address space of the \ac{cpu}.
If the \ac{cpu} has separate address spaces for \ac{mmio} and \ac{io} (e.g., for x86 \acp{cpu}), a \ac{pci}(e) card can have memory regions in both address spaces.
For this reason, the \ac{pci} host bridge has three different \ac{tlm} target sockets to allow different address spaces for the configuration, \ac{mmio}, and \ac{io} spaces.
In \cref{fig:bar}, we assume a single address space.
The \ac{cpu} read and write transactions to the memory-mapped regions of the \ac{pci} device are received by the \ac{pci} host bridge.
The bridge knows the mappings of the different devices and translates the \ac{tlm} transaction into a \ac{pci} transaction using \ac{vcml}'s \ac{pci} \ac{tlm} protocol.

To process the request, the \ac{vpci} device needs to access the \ac{hw} device.
For this access, two possibilities exist.
If the address belongs to the configuration space, Linux's \textit{pread} or \textit{pwrite} syscalls need to be used.
The \acl{fd} that needs to be passed as a parameter is the one of the device's \ac{vfio} group.
For accesses to regions of the \ac{mmio} and \ac{io} spaces, the memory region of the device can be mapped into the \ac{vp}'s virtual address space.
This allows access without further \ac{vfio} involvement.
The configuration space cannot be directly mapped because some fields are virtualized and therefore need the involvement of \ac{vfio}.

\subsection{\acf{dma}}
Most \ac{pci}(e) devices use \ac{dma} to directly access memory.
The \ac{sw} running on the \ac{cpu} configures the \ac{pci}(e) device by communicating \acp{pa} that contain data to be processed.
This is also done by the \ac{sw} running on the \ac{vcpu}.
Since a \ac{vp} is a user-space process running on a host machine, \acp{pa} from the target \ac{sw}'s point of view are not real physical addresses of the host machine.
When the target \ac{sw} on the \ac{vp} configures the \ac{pci}(e) device, an address translation is needed.

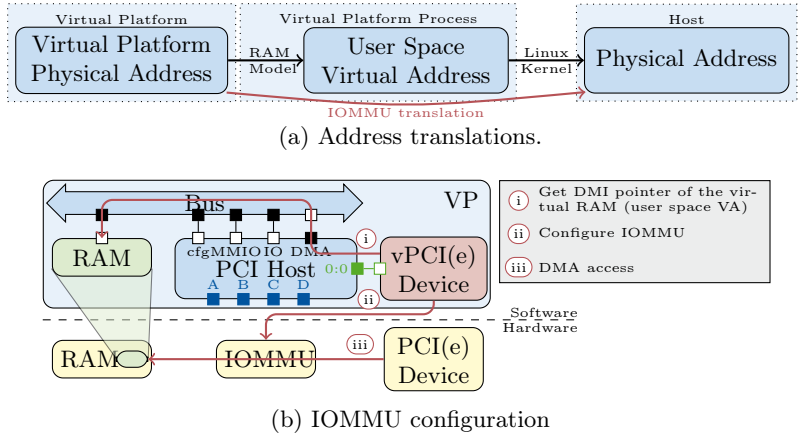
\begin{figure}[!t]
    \centering
    \begin{subfigure}[B]{\linewidth}
        \centering
        \tikzstyle{nodetype} = [rectangle, rounded corners, minimum width=0.8cm, minimum height=0.87cm, text centered, text width=2.0cm, draw=black, fill=rwth-25, font={\footnotesize}]
\tikzstyle{isock}    = [rectangle, minimum width=.15cm, minimum height=.15cm, draw=black, fill=black, inner sep=0pt]
\tikzstyle{tsock}    = [isock, fill=white]
\tikzstyle{isocki}   = [isock, fill=rwth, draw=rwth]
\tikzstyle{tsocki}   = [isocki, fill=white]
\tikzstyle{isockp}   = [isock, fill=grun, draw=grun]
\tikzstyle{tsockp}   = [isockp, fill=white]
\tikzstyle{lbl}      = [font={\tiny}, inner sep=1pt]
\tikzstyle{socklbl}  = [lbl]

\pgfdeclarelayer{background}
\pgfsetlayers{background,main}

\begin{tikzpicture}[node distance=.5cm and .5cm]
    \begin{scope}[name prefix=userspace-]
        \node[nodetype, text width=2.5cm] (main) {User Space \acl{va}};
    \end{scope}

    \begin{scope}[name prefix=vp-]
        \node[nodetype, anchor=east, text width=2.6cm] at ([xshift=-1.0cm]userspace-main.west) (main) {\acl{vp} \acl{pa}};
        \node[lbl, anchor=south] at (main.north) (lbl) {\acl{vp}};
        \begin{pgfonlayer}{background}
            \node[fit={(main) (lbl)}, draw=black, dotted, inner sep=0.1cm, fill=rwth-10] {};
        \end{pgfonlayer}
    \end{scope}

    \begin{scope}[name prefix=pa-]
        \node[nodetype, anchor=west, text width=2.5cm] at ([xshift=+1.0cm]userspace-main.east) (main) {\acl{pa}};
        \node[lbl, anchor=south] at (main.north) (lbl) {Host};
        \begin{pgfonlayer}{background}
            \node[fit={(main) (lbl)}, draw=black, dotted, inner sep=0.1cm, fill=rwth-10] {};
        \end{pgfonlayer}
    \end{scope}

    \draw[->, thick] (vp-main.east) -- node[lbl, text width=0.45cm, anchor=center, text centered, inner sep=0cm] (ram-lbl) {\acs{ram} Model} (userspace-main.west);
    \draw[->, thick] (userspace-main.east) -- node[lbl, text width=0.8cm, anchor=center, text centered] {Linux Kernel} (pa-main.west);

    \begin{scope}[name prefix=userspace-]
        \phantom{\node[fit={(ram-lbl) (main)}, inner sep=0pt] (bb) {};}
        \node[lbl, anchor=south] at (bb.north) (lbl) {\acl{vp} Process};
        \begin{pgfonlayer}{background}
            \node[fit={(main) (lbl) (ram-lbl)}, draw=black, dotted, inner sep=0.1cm, fill=rwth-10] {};
        \end{pgfonlayer}
    \end{scope}

    \draw[->, thick, bordeaux-75] ([xshift=-0.02cm, yshift=+0.02cm]vp-main.south east) .. controls ([yshift=-0.2cm]userspace-main.south) .. node[lbl, below] {\acs{iommu} translation} ([xshift=+0.02cm, yshift=+0.02cm]pa-main.south west);

\end{tikzpicture}
        \vspace{-0.5em}
        \caption{Address translations.}
        \label{fig:address}
        \vspace{1em}
    \end{subfigure}
    \begin{subfigure}[B]{\linewidth}
        \centering
        \tikzstyle{nodetype} = [rectangle, rounded corners, minimum width=1.1cm, minimum height=.5cm, text centered, text width=1.1cm, draw=black, fill=rwth-25, font={\footnotesize}]
\tikzstyle{isock}    = [rectangle, minimum width=.15cm, minimum height=.15cm, draw=black, fill=black, inner sep=0pt]
\tikzstyle{tsock}    = [isock, fill=white]
\tikzstyle{isocki}   = [isock, fill=rwth, draw=rwth]
\tikzstyle{tsocki}   = [isocki, fill=white]
\tikzstyle{isockp}   = [isock, fill=grun, draw=grun]
\tikzstyle{tsockp}   = [isockp, fill=white]
\tikzstyle{lbl}      = [font={\tiny}, inner sep=1pt]
\tikzstyle{socklbl}  = [lbl]

\pgfdeclarelayer{background}
\pgfdeclarelayer{middle}
\pgfsetlayers{background,middle,main}

\begin{tikzpicture}[node distance=.5cm and .5cm]
    \begin{scope}[name prefix=vpram-]
        \node[nodetype, fill=grun-25] (main) {RAM};
        \node[tsock] at (main.north) (sock) {};
    \end{scope}

    \begin{scope}[name prefix=pci-host-]
        \node[nodetype, minimum width=1.9cm, minimum height=0.8cm, anchor=north west, text width=2.2cm] at ([xshift=+0.3cm]vpram-main.north east) (main) {PCI Host};

        \node[tsock, xshift=-0.9cm] at (main.north) (sock_cfg) {};
        \node[socklbl, anchor=north] at (sock_cfg.south) {cfg};

        \node[tsock, xshift=-0.4cm] at (main.north) (sock_mmio) {};
        \node[socklbl, anchor=north] at (sock_mmio.south) {MMIO};

        \node[tsock, xshift=+0.1cm] at (main.north) (sock_io) {};
        \node[socklbl, anchor=north] at (sock_io.south) {IO};

        \node[isock, xshift=+0.6cm] at (main.north) (sock_dma) {};
        \node[socklbl, anchor=north] at (sock_dma.south) {DMA};

        \node[isocki, xshift=-0.7cm] at (main.south) (sock_inta) {};
        \node[socklbl, anchor=base, text=rwth] at ([yshift=0.05cm]sock_inta.north) {A};

        \node[isocki, xshift=-0.3cm] at (main.south) (sock_intb) {};
        \node[socklbl, anchor=base, text=rwth] at ([yshift=0.05cm]sock_intb.north) {B};

        \node[isocki, xshift=+0.1cm] at (main.south) (sock_intc) {};
        \node[socklbl, anchor=base, text=rwth] at ([yshift=0.05cm]sock_intc.north) {C};

        \node[isocki, xshift=+0.5cm] at (main.south) (sock_intd) {};
        \node[socklbl, anchor=base, text=rwth] at ([yshift=0.05cm]sock_intd.north) {D};

        \node[isockp] at (main.east) (sock_pci0_0) {};
        \node[socklbl, anchor=east, text=grun] at (sock_pci0_0.west) {0:0};
    \end{scope}

    \begin{scope}[name prefix=vfio-]
        \node[nodetype, fill=bordeaux-25, minimum height=0.8cm, anchor=west, text width=1.2cm] at ([xshift=0.3cm]pci-host-main.east) (main) {\acs{vpci}(e) Device};
        \node[tsockp] at (main.west) (sock) {};
    \end{scope}

    \node[lbl, fit={(pci-host-main) (vpram-main)}, inner sep=0pt] (peripherals) {};

    \begin{scope}[name prefix=bus-]
        \node[double arrow, draw, minimum height=4.2cm, fill=rwth-25, font={\footnotesize}, anchor=south, inner sep=1pt, double arrow head extend=0.1cm] at ([yshift=+0.3cm]peripherals.north) (main) {Bus};

        \node[isock] at (main.south-|pci-host-sock_cfg) (pci-host-cfg){};
        \node[isock] at (main.south-|pci-host-sock_mmio) (pci-host-mmio){};
        \node[isock] at (main.south-|pci-host-sock_io) (pci-host-io){};
        \node[tsock] at (main.south-|pci-host-sock_dma) (pci-host-dma){};
        \draw (pci-host-cfg) -- (pci-host-sock_cfg);
        \draw (pci-host-mmio) -- (pci-host-sock_mmio);
        \draw (pci-host-io) -- (pci-host-sock_io);
        \draw (pci-host-dma) -- (pci-host-sock_dma);

        \node[isock] at (main.south-|vpram-sock) (vpram){};
        \draw (vpram) -- (vpram-sock);

    \end{scope}

    \draw[draw=grun] (vfio-sock) -- (pci-host-sock_pci0_0);

    \node[anchor=north east] at (vfio-main.east|-bus-main.after tip 2) (lbl_vp) {VP};
    \begin{pgfonlayer}{background}
        \node[nodetype, fill=rwth-10, inner sep=1pt, fit={(vfio-main) (lbl_vp) (pci-host-main) (pci-host-sock_inta) (bus-main) (bus-main.after tip 2)}] (vp) {};
    \end{pgfonlayer}

    \begin{scope}[name prefix=pci-]
        \node[nodetype, fill=yellow-25, anchor=north] at ([yshift=-0.35cm]vfio-main.south) (main) {PCI(e) Device};
    \end{scope}

    \begin{scope}[name prefix=ram-]
        \begin{pgfonlayer}{background}
            \node[nodetype, fill=yellow-25, anchor=center, align=left] at (vpram-main.center|-pci-main.center) (main) {RAM};
        \end{pgfonlayer}

        \node[nodetype, fill=grun-25, anchor=east, minimum width=0.2cm, text width=0.2cm, minimum height=0.1cm] (vp-part) at ([xshift=-0.05cm]main.east) {};
        \begin{pgfonlayer}{middle}
            \draw[fill=grun-25, semitransparent] ([yshift=-0.1cm]vp-part.north east) -- ([yshift=+0.1cm]vpram-main.south east) -- ([yshift=+0.1cm]vpram-main.south west) -- ([yshift=-0.1cm]vp-part.north west) -- ([yshift=-0.1cm]vp-part.north east);
        \end{pgfonlayer}
    \end{scope}

    \begin{scope}[name prefix=iommu-]
        \node[nodetype, fill=yellow-25, anchor=center] at (pci-main.center-|pci-host-main) (main) {IOMMU};
    \end{scope}

    \begin{pgfonlayer}{background}
        \draw[dashed] ([yshift=-0.15cm]vp.south west) -- ([yshift=-0.15cm]vp.south east) -- +(1.2cm, 0cm)
        node[lbl, anchor=south east] {Software}
        node[lbl, anchor=north east] {Hardware};
    \end{pgfonlayer}

    \draw[->, bordeaux-75, rounded corners, thick] ([yshift=0.2cm]vfio-main.west) -|  node[lbl, pos=0.1, anchor=south, yshift=0.02cm] {\Circled{i}}  ([yshift=0.2cm]bus-pci-host-dma.center) -| (vpram-sock.center);
    \draw[->, bordeaux-75, rounded corners, thick] (vfio-main.south) |-  node[lbl, pos=1, anchor=south, yshift=0.02cm] {\Circled{ii}}  ([xshift=0.15cm, yshift=-0.2cm]pci-host-main.south east) -| (iommu-main.north);
    \draw[->, bordeaux-75, rounded corners, thick] (pci-main.west) -- node[lbl, pos=0.1, anchor=south, yshift=0.02cm] {\Circled{iii}} (ram-vp-part.east);

    \node[lbl, anchor=north west] at ([xshift=0.2cm, yshift=-3pt]vp.north east) (no1) {\Circled{i}};
    \node[lbl, anchor=west, text width=3.0cm] at ([xshift=0.05cm]no1.east) (no1_lbl) {Get DMI pointer of the virtual RAM (user space \ac{va})};

    \node[lbl, anchor=north] at ([yshift=-0.1cm]no1.south) (no2) {\Circled{ii}};
    \node[lbl, anchor=west] at (no2.east-|no1_lbl.west) (no2_lbl) {Configure IOMMU};

    \node[lbl, anchor=north] at ([yshift=-0.1cm]no2.south) (no3) {\Circled{iii}};
    \node[lbl, anchor=west] at (no3.east-|no1_lbl.west) (no3_lbl) {DMA access};

    \begin{pgfonlayer}{background}
        \node[draw=black, fill=black-10, inner sep=1pt, fit={{(no1) (no2) (no3) (no1_lbl) (no2_lbl) (no3_lbl)}}] (legend) {};
    \end{pgfonlayer}

\end{tikzpicture}
        \caption{\acs{iommu} configuration}
        \label{fig:dma}
    \end{subfigure}
    \caption{\acs{dma} access configuration.}
    \vspace{-2em}
\end{figure}

The performed address translations are shown in \cref{fig:address}.
When the \ac{sw} running on the \ac{vcpu} communicates addresses to the \ac{vpci} device, those addresses are within the physical address space of the \ac{vp}.
They are referring to the \ac{ram} model of the \ac{vp} which is somewhere placed in the actual \ac{ram} of the host machine.
The \ac{ram} model of the \ac{vp} can translate a \ac{pa} of the \ac{vp} into a \ac{va} of the \ac{vp} process on the host machine.
This \ac{va} can be translated into a host \ac{pa} by the host's Linux kernel.
The \ac{iommu} of the host is configured to perform the same mapping.
Once the mapping has been configured, the \ac{vcpu} can communicate \ac{vp} \acp{pa} to the \ac{pci}(e) device and the \ac{iommu} maps them to the corresponding host \acp{pa} before a \ac{dma} access.

\Cref{fig:dma} shows the configuration of host's \ac{iommu} in more detail.
The user of the \ac{vp} needs to specify the physical \ac{vp} address range the \ac{pci}(e) should be able to access using \ac{dma}.
This is usually the address range of the \ac{vp}'s \ac{ram}.
During the \ac{vp} setup, the \ac{vpci} device requests a \ac{dmi} pointer (host \ac{va}) to the specified region using the \ac{dmi} \ac{tlm} socket of the connected \ac{pci} host bridge~\Circled{i}.
The \ac{vpci} device configures the host's \ac{iommu} by passing the received \ac{dmi} pointer (host \ac{va}) and the corresponding \ac{vp}'s \ac{pa} range to a \ac{vfio} system call~\Circled{ii}.
After the \ac{iommu} has been configured, the \ac{vcpu} can communicate \ac{vp} \acp{pa} to the \ac{pci} device and the \ac{iommu} performs the mapping to host \acp{pa}~\Circled{iii}.

\subsection{Interrupt Handling}

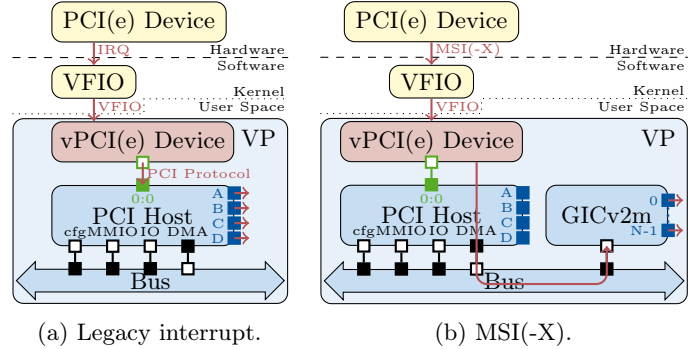
\begin{figure}[!t]
    \centering

    \begin{subfigure}[c]{0.3\linewidth}
        \centering
        \tikzstyle{nodetype} = [rectangle, rounded corners, minimum width=1.1cm, minimum height=.4cm, text centered, text width=1.1cm, draw=black, fill=rwth-25, font={\footnotesize}]
\tikzstyle{isock}    = [rectangle, thick, minimum width=.15cm, minimum height=.15cm, draw=black, fill=black, inner sep=0pt]
\tikzstyle{tsock}    = [isock, fill=white]
\tikzstyle{isocki}   = [isock, fill=rwth, draw=rwth]
\tikzstyle{tsocki}   = [isocki, fill=white]
\tikzstyle{isockp}   = [isock, fill=grun, draw=grun]
\tikzstyle{tsockp}   = [isockp, fill=white]
\tikzstyle{lbl}      = [font={\tiny}, inner sep=1pt]
\tikzstyle{socklbl}  = [lbl]

\pgfdeclarelayer{background}
\pgfsetlayers{background,main}

\begin{tikzpicture}[node distance=.5cm and .5cm]
    \begin{scope}[name prefix=pci-]
        \node[nodetype, fill=yellow-25, text width=2.1cm] (main) {PCI(e) Device};
    \end{scope}

    \begin{scope}[name prefix=linux-]
        \node[nodetype, fill=yellow-25, text width=0.9cm, anchor=north west] (main) at ([yshift=-0.3cm]pci-main.south west) {VFIO};
    \end{scope}

    \begin{scope}[name prefix=vfio-]
        \node[nodetype, fill=bordeaux-25, anchor=north west, text width=2.2cm] (main) at ([yshift=-0.3cm]linux-main.south west) {\acs{vpci}(e) Device};
        \node[tsockp] at (main.south) (sock) {};
    \end{scope}

    \begin{scope}[name prefix=pci-host-]
        \node[nodetype, minimum width=1.9cm, minimum height=0.8cm, anchor=north west, text width=2.2cm] at ([yshift=-0.3cm]vfio-main.south west) (main) {PCI Host};

        \node[tsock, xshift=-0.9cm] at (main.south) (sock_cfg) {};
        \node[socklbl, anchor=base] at ([yshift=0.05cm]sock_cfg.north) {cfg};

        \node[tsock, xshift=-0.4cm] at (main.south) (sock_mmio) {};
        \node[socklbl, anchor=base] at ([yshift=0.05cm]sock_mmio.north) {MMIO};

        \node[tsock, xshift=+0.1cm] at (main.south) (sock_io) {};
        \node[socklbl, anchor=base] at ([yshift=0.05cm]sock_io.north) {IO};

        \node[isock, xshift=+0.6cm] at (main.south) (sock_dma) {};
        \node[socklbl, anchor=base] at ([yshift=0.05cm]sock_dma.north) {DMA};

        \node[isocki, yshift=+0.3cm] at (main.east) (sock_inta) {};
        \node[socklbl, anchor=east, text=rwth] at (sock_inta.west) {A};

        \node[isocki, yshift=+0.1cm] at (main.east) (sock_intb) {};
        \node[socklbl, anchor=east, text=rwth] at (sock_intb.west) {B};

        \node[isocki, yshift=-0.1cm] at (main.east) (sock_intc) {};
        \node[socklbl, anchor=east, text=rwth] at (sock_intc.west) {C};

        \node[isocki, yshift=-0.3cm] at (main.east) (sock_intd) {};
        \node[socklbl, anchor=east, text=rwth] at (sock_intd.west) {D};

        \node[isockp] at (main.north-|vfio-sock) (sock_pci0_0) {};
        \node[socklbl, anchor=north, text=grun] at (sock_pci0_0.south) {0:0};
    \end{scope}

    \begin{scope}[name prefix=bus-]
        \node[double arrow, draw, minimum height=\linewidth-0.1cm, fill=rwth-25, font={\footnotesize}, anchor=north, inner sep=1pt, double arrow head extend=0.1cm] at ([xshift=0.1cm, yshift=-0.3cm]pci-host-main.south) (main) {Bus};

        \node[isock] at (main.north-|pci-host-sock_cfg) (pci-host-cfg){};
        \node[isock] at (main.north-|pci-host-sock_mmio) (pci-host-mmio){};
        \node[isock] at (main.north-|pci-host-sock_io) (pci-host-io){};
        \node[tsock] at (main.north-|pci-host-sock_dma) (pci-host-dma){};
        \draw[thick] (pci-host-cfg) -- (pci-host-sock_cfg);
        \draw[thick] (pci-host-mmio) -- (pci-host-sock_mmio);
        \draw[thick] (pci-host-io) -- (pci-host-sock_io);
        \draw[thick] (pci-host-dma) -- (pci-host-sock_dma);
    \end{scope}

    \draw[draw=grun] (vfio-sock) -- (pci-host-sock_pci0_0);

    \draw[->, bordeaux-75, thick] (pci-main.south-|linux-main.center) -- node[lbl, anchor=west, pos=0.35] {IRQ} (linux-main);
    \draw[->, bordeaux-75, thick] (linux-main) -- node[lbl, anchor=west, pos=0.35] {VFIO} (vfio-main.north-|linux-main.center);
    \draw[->, bordeaux-75, thick] (vfio-sock.center) -- node[lbl, anchor=west] {PCI Protocol} (pci-host-sock_pci0_0.center);
    \draw[->, bordeaux-75, thick] (pci-host-sock_inta.center) -- +(0.2cm,0cm) node[lbl, anchor=west] (lbl-sig-a){};
    \draw[->, bordeaux-75, thick] (pci-host-sock_intb.center) -- +(0.2cm,0cm) node[lbl, anchor=west] (lbl-sig-b){};
    \draw[->, bordeaux-75, thick] (pci-host-sock_intc.center) -- +(0.2cm,0cm) node[lbl, anchor=west] (lbl-sig-c){};
    \draw[->, bordeaux-75, thick] (pci-host-sock_intd.center) -- +(0.2cm,0cm) node[lbl, anchor=west] (lbl-sig-d){};

    \node[anchor=north east] at (vfio-main.north east-|bus-main.east) (lbl_vp) {VP};
    \begin{pgfonlayer}{background}
        \node[nodetype, fill=rwth-10, inner sep=1pt, fit={(vfio-main) (lbl_vp) (pci-host-main) (pci-host-sock_cfg) (bus-main) (bus-main.after tip 1)}] (vp) {};

        \draw[dotted] ([yshift=-0.2cm]linux-main.south-|vp.west)
            -- ([xshift=0.1cm,yshift=-0.2cm]linux-main.south east) node (help) {}
            -- (help|-linux-main.south east)
            -- (linux-main.south east-|vp.east)
                node[lbl,anchor=south east] {Kernel}
                node[lbl,anchor=north east] {User Space} ;

        \draw[dashed] ([yshift=-0.2cm]pci-main.south-|vp.west)
            -- ([xshift=0.1cm,yshift=-0.2cm]pci-main.south east)
            -- ([yshift=-0.2cm]pci-main.south-|vp.east)
                node[lbl,anchor=south east] {Hardware}
                node[lbl,anchor=north east] {Software} ;
    \end{pgfonlayer}

\end{tikzpicture}%
        \caption{Legacy interrupt.}
        \label{fig:irq:legacy}
    \end{subfigure}
    \quad
    \begin{subfigure}[c]{0.4\linewidth}
        \centering
        \tikzstyle{nodetype} = [rectangle, rounded corners, minimum width=1.1cm, minimum height=.4cm, text centered, text width=1.1cm, draw=black, fill=rwth-25, font={\footnotesize}]
\tikzstyle{isock}    = [rectangle, thick, minimum width=.15cm, minimum height=.15cm, draw=black, fill=black, inner sep=0pt]
\tikzstyle{tsock}    = [isock, fill=white]
\tikzstyle{isocki}   = [isock, fill=rwth, draw=rwth]
\tikzstyle{tsocki}   = [isocki, fill=white]
\tikzstyle{isockp}   = [isock, fill=grun, draw=grun]
\tikzstyle{tsockp}   = [isockp, fill=white]
\tikzstyle{lbl}      = [font={\tiny}, inner sep=1pt]
\tikzstyle{socklbl}  = [lbl]

\pgfdeclarelayer{background}
\pgfsetlayers{background,main}

\begin{tikzpicture}[node distance=.5cm and .5cm]
    \begin{scope}[name prefix=pci-]
        \node[nodetype, fill=yellow-25, text width=2.1cm] (main) {PCI(e) Device};
    \end{scope}

    \begin{scope}[name prefix=linux-]
        \node[nodetype, fill=yellow-25, text width=0.9cm, anchor=north] (main) at ([yshift=-0.3cm]pci-main.south) {VFIO};
    \end{scope}

    \begin{scope}[name prefix=vfio-]
        \node[nodetype, fill=bordeaux-25, anchor=north, text width=2.2cm] (main) at ([yshift=-0.3cm]linux-main.south) {\acs{vpci}(e) Device};
        \node[tsockp] at (main.south) (sock) {};
    \end{scope}

    \begin{scope}[name prefix=pci-host-]
        \node[nodetype, minimum width=1.9cm, minimum height=0.8cm, anchor=north, text width=2.2cm] at ([yshift=-0.3cm]vfio-main.south) (main) {PCI Host};

        \node[tsock, xshift=-0.9cm] at (main.south) (sock_cfg) {};
        \node[socklbl, anchor=base] at ([yshift=0.05cm]sock_cfg.north) {cfg};

        \node[tsock, xshift=-0.4cm] at (main.south) (sock_mmio) {};
        \node[socklbl, anchor=base] at ([yshift=0.05cm]sock_mmio.north) {MMIO};

        \node[tsock, xshift=+0.1cm] at (main.south) (sock_io) {};
        \node[socklbl, anchor=base] at ([yshift=0.05cm]sock_io.north) {IO};

        \node[isock, xshift=+0.6cm] at (main.south) (sock_dma) {};
        \node[socklbl, anchor=base] at ([yshift=0.05cm]sock_dma.north) {DMA};

        \node[isocki, yshift=+0.3cm] at (main.east) (sock_inta) {};
        \node[socklbl, anchor=east, text=rwth] at (sock_inta.west) {A};

        \node[isocki, yshift=+0.1cm] at (main.east) (sock_intb) {};
        \node[socklbl, anchor=east, text=rwth] at (sock_intb.west) {B};

        \node[isocki, yshift=-0.1cm] at (main.east) (sock_intc) {};
        \node[socklbl, anchor=east, text=rwth] at (sock_intc.west) {C};

        \node[isocki, yshift=-0.3cm] at (main.east) (sock_intd) {};
        \node[socklbl, anchor=east, text=rwth] at (sock_intd.west) {D};

        \node[isockp] at (main.north-|vfio-sock) (sock_pci0_0) {};
        \node[socklbl, anchor=north, text=grun] at (sock_pci0_0.south) {0:0};
    \end{scope}

    \begin{scope}[name prefix=gicv2m-]
        \node[nodetype, minimum height=0.8cm, anchor=west, text width=1.4cm] at ([xshift=0.3cm]pci-host-main.east) (main) {GICv2m};

        \node[tsock, xshift=+0.0cm] at (main.south) (sock) {};

        \node[isocki, yshift=+0.2cm] at (main.east) (sock_int0) {};
        \node[socklbl, anchor=east, text=rwth] at (sock_int0.west) {0};

        \node[yshift=+0.0cm, text=rwth, rotate=90] at ([xshift=-0.00cm]main.east) (sock_int1) {...};

        \node[isocki, yshift=-0.2cm] at (main.east) (sock_intn) {};
        \node[socklbl, anchor=east, text=rwth] at (sock_intn.west) {N-1};
    \end{scope}

    \node[lbl, fit={(pci-host-main) (gicv2m-main)}, inner sep=0pt] (peripherals) {};

    \begin{scope}[name prefix=bus-]
        \node[double arrow, draw, minimum height=\linewidth-0.1cm, fill=rwth-25, font={\footnotesize}, anchor=north, inner sep=1pt, double arrow head extend=0.1cm] at ([yshift=-0.3cm]peripherals.south) (main) {Bus};

        \node[isock] at (main.north-|pci-host-sock_cfg) (pci-host-cfg){};
        \node[isock] at (main.north-|pci-host-sock_mmio) (pci-host-mmio){};
        \node[isock] at (main.north-|pci-host-sock_io) (pci-host-io){};
        \node[tsock] at (main.north-|pci-host-sock_dma) (pci-host-dma){};
        \draw[thick] (pci-host-cfg) -- (pci-host-sock_cfg);
        \draw[thick] (pci-host-mmio) -- (pci-host-sock_mmio);
        \draw[thick] (pci-host-io) -- (pci-host-sock_io);
        \draw[thick] (pci-host-dma) -- (pci-host-sock_dma);

        \node[isock] at (main.north-|gicv2m-sock) (gicv2m){};
        \draw[thick] (gicv2m) -- (gicv2m-sock);
    \end{scope}

    \draw[draw=grun, thick] (vfio-sock) -- (pci-host-sock_pci0_0);

    \draw[->, bordeaux-75, thick] (pci-main) -- node[lbl, anchor=west, pos=0.35] {MSI(-X)} (linux-main);
    \draw[->, bordeaux-75, thick] (linux-main) -- node[lbl, anchor=west, pos=0.35] {VFIO} (vfio-main);
    \draw[->, bordeaux-75, thick, rounded corners] (vfio-main.south-|bus-pci-host-dma) -- ([yshift=-0.2cm]bus-pci-host-dma.center) -| (gicv2m-main.south);
    \draw[->, bordeaux-75, thick] (gicv2m-sock_int0.center) -- +(0.2cm,0cm) node[lbl, anchor=west] (lbl-sig-a){};
    \draw[->, bordeaux-75, thick] (gicv2m-sock_intn.center) -- +(0.2cm,0cm) node[lbl, anchor=west] (lbl-sig-a){};

    \node[anchor=north east] at (vfio-main.north east-|bus-main.east) (lbl_vp) {VP};
    \begin{pgfonlayer}{background}
        \node[nodetype, fill=rwth-10, inner sep=1pt, fit={(vfio-main) (lbl_vp) (pci-host-main) (pci-host-sock_cfg) (bus-main) (bus-main.after tip 1)}] (vp) {};

        \draw[dotted] ([yshift=-0.2cm]linux-main.south west-|vp.west)
            -- ([xshift=0.1cm,yshift=-0.2cm]linux-main.south east) node (help) {}
            -- (help|-linux-main.south east)
            -- (linux-main.south east-|vp.north east)
                node[lbl,anchor=south east] {Kernel}
                node[lbl,anchor=north east] {User Space} ;

        \draw[dashed] ([yshift=-0.2cm]pci-main.south west-|vp.west)
            -- ([xshift=0.1cm,yshift=-0.2cm]pci-main.south east)
            -- ([yshift=-0.2cm]pci-main.south west-|vp.east)
                node[lbl,anchor=south east] {Hardware}
                node[lbl,anchor=north east] {Software} ;
    \end{pgfonlayer}

\end{tikzpicture}%
        \caption{\acs{msi}(-X).}
        \label{fig:irq:msi}
    \end{subfigure}

    \caption{Interrupt forwarding implementation.}
    \label{fig:irq}
    \vspace{-2em}
\end{figure}

When the physical \ac{pci}(e) card sends an interrupt, the interrupt is processed by the \ac{vfio} driver.
The interrupt implementation is visualized in \cref{fig:irq}.
\Cref{fig:irq:legacy} shows the forwarding of legacy interrupts.
When the \ac{pci}(e) device sends a legacy interrupt, the \ac{vpci} device reads the \ac{irq} pin number (\textit{A}, \textit{B}, \textit{C}, or \textit{D}) from the configuration region.
It then sends an \ac{irq} containing the pin number to the \ac{pci} host using the backward path of the \ac{vpci}'s \ac{pci} \ac{tlm} socket.
The \ac{pci} host bridge signals the \ac{irq} to the connected interrupt controller using \ac{vcml}'s \ac{gpio} protocol.

In case the interrupt is an \ac{msi}(-X), the \ac{vpci} device needs to perform a write operation to the corresponding \ac{msi}(-X) region (see \cref{fig:irq:msi}).
This region is usually part of the \ac{msi}(-X)-capable interrupt controller, such as the \textit{GICv2m} extension.
The destination address of the write transaction is configured by the \ac{os} running on the \ac{vcpu}.
For \acp{msi}, the target address is written in a dedicated field in the configuration space of the device.
This field is virtualized by \ac{vfio} meaning a read to this field using \ac{vfio}'s \ac{api} does not return the value of the device.
Thereby, different \ac{msi} addresses for the \ac{vpci}(e) and real \ac{pci}(e) devices are possible.

In the case of a \ac{msi}-X, the destination addresses are written in a table that is placed in the \ac{mmio} region of the device by the \ac{os}.
After the \ac{vpci} device has read the destination address of the \ac{msi}(-X), a transaction is sent through the \ac{dma} \ac{tlm} socket of the \ac{pci} host bridge.
The \ac{msi}(-X) capable interrupt controller then signals an \ac{irq}.

\section{Case Study}
\label{sec:casestudy}

\begin{figure}[!t]
    \centering
    \input{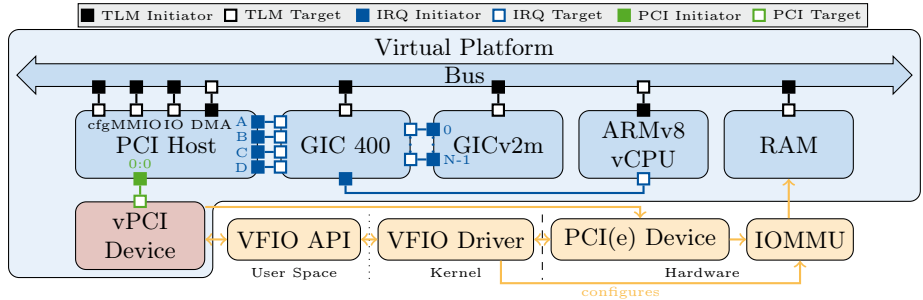}
    \vspace{-2em}
    \caption{\acs{vp} setup with connected \acs{vfio} device.}
    \label{fig:vp}
    \vspace{-1em}
\end{figure}

To showcase the effectiveness and working principle of our \ac{vpci} model, we present the integration of Google Coral's \ac{pcie} Edge \ac{tpu} into the \ac{avp64}~\cite{junger_armv8_2023,junger_fast_2019}.
We compare the runtime performance of an \ac{ai}-workload execution on the \ac{vp} with and without \ac{tpu} offloading.
The workloads are based on \ac{tfl}, which is designed for embedded and edge devices~\cite{martin_abadi_tensorflow_2015}.
\Cref{tab:benchmark} shows the used benchmarks and parameters.

\begin{table}[!b]
    \vspace{-1em}
    \centering
    \caption{Benchmark parameters (batch size = \num{1}).}
    \vspace{1em}
    \label{tab:benchmark}
    \footnotesize
    \begin{tabular}{cc||c|r|c}
        \textbf{}                                                 & \textbf{Neural Network}                                                                   & \textbf{Input Size}     & \textbf{Model Size}   & \textbf{Dataset}                                                                       \\ \hline \hline
        \multirow{7}{*}{\rotatebox[origin=c]{90}{Classification}} & \textbf{EfficientNet L}~\cite{tan_efficientnet_2019}                                      & 300$\times$300$\times$3 & \SI{12.8}{\mega\byte} & \multirow{7}{*}{\rotatebox[origin=c]{90}{ILSVRC2012~\cite{russakovsky_imagenet_2015}}} \\ 
                                                                  & \textbf{EfficientNet M}~\cite{tan_efficientnet_2019}                                      & 240$\times$240$\times$3 & \SI{8.7}{\mega\byte}  &                                                                                        \\ 
                                                                  & \textbf{EfficientNet S}~\cite{tan_efficientnet_2019}                                      & 224$\times$224$\times$3 & \SI{6.8}{\mega\byte}  &                                                                                        \\ 
                                                                  & \textbf{MobileNet V1}~\cite{howard_mobilenets_2017}                                       & 224$\times$224$\times$3 & \SI{4.5}{\mega\byte}  &                                                                                        \\ 
                                                                  & \textbf{MobileNet V2}~\cite{sandler_mobilenetv2_2018}                                     & 224$\times$224$\times$3 & \SI{4.1}{\mega\byte}  &                                                                                        \\ 
                                                                  & \textbf{MobileNet V3}~\cite{howard_searching_2019}                                        & 224$\times$224$\times$3 & \SI{4.9}{\mega\byte}  &                                                                                        \\ 
                                                                  & \textbf{ResNet-50}~\cite{he_deep_2016}                                                    & 224$\times$224$\times$3 & \SI{25.0}{\mega\byte} &                                                                                        \\ \hline
        \multirow{7}{*}{\rotatebox[origin=c]{90}{Detection}}      & \textbf{EfficientDet-Lite 0}~\cite{tan_efficientdet_2020}                                 & 320$\times$320$\times$3 & \SI{5.7}{\mega\byte}  & \multirow{7}{*}{\rotatebox[origin=c]{90}{COCO~\cite{lin_microsoft_2014}}}              \\ 
                                                                  & \textbf{EfficientDet-Lite 1}~\cite{tan_efficientdet_2020}                                 & 384$\times$384$\times$3 & \SI{7.6}{\mega\byte}  &                                                                                        \\ 
                                                                  & \textbf{EfficientDet-Lite 2}~\cite{tan_efficientdet_2020}                                 & 448$\times$448$\times$3 & \SI{10.2}{\mega\byte} &                                                                                        \\ 
                                                                  & \textbf{EfficientDet-Lite 3}~\cite{tan_efficientdet_2020}                                 & 512$\times$512$\times$3 & \SI{14.4}{\mega\byte} &                                                                                        \\ 
                                                                  & \textbf{EfficientDet-Lite 3x}~\cite{tan_efficientdet_2020}                                & 640$\times$640$\times$3 & \SI{20.6}{\mega\byte} &                                                                                        \\ 
                                                                  & \textbf{SSD MobileNet V2}~\cite{chiu_mobilenet-ssdv2_2020}                                & 300$\times$300$\times$3 & \SI{6.7}{\mega\byte}  &                                                                                        \\ 
                                                                  & \textbf{SSD/FPN MobileNet V1}~\cite{howard_mobilenets_2017,lin_feature_2017,liu_ssd_2016} & 640$\times$640$\times$3 & \SI{7.0}{\mega\byte}  &                                                                                        \\
    \end{tabular}
\end{table}

To connect the \ac{vpci} device to \ac{avp64}, we have to extend the \ac{vp} by a \ac{pci} host bridge and a \textit{\ac{gic}v2m}, which adds \ac{msi}(-X) capabilities to the \textit{\ac{gic}~400}.
Both peripherals are taken from \ac{vcml}~\cite{machineware_machineware-gmbhvcml_2024}.
The full \ac{vp} setup is shown in \cref{fig:vp}.
The \ac{vpci} device is connected to the \ac{pci} Host using \ac{vcml}'s \ac{tlm}-based \ac{pci} sockets.
The \ac{pci} Host is connected to the system bus, which allows configuration from the \ac{cpu} and access to the \ac{ram}.
We use Buildroot~\cite{buildrootorg_buildrootorg_2024} to create a Linux image that contains the driver for the \ac{tpu}, Python, and the required packages including \ac{tfl}.

\subsection{Speedup Analysis}

To analyze the impact of offloading workload to the \ac{tpu}, all benchmarks are executed in two configurations.
At first, they are compiled for \ac{cpu}-only execution.
Then, they are compiled using the Edge \ac{tpu} compiler to offload computation to the \ac{tpu}.
The executed workloads perform image classification and detection tasks using different \acp{nn}.

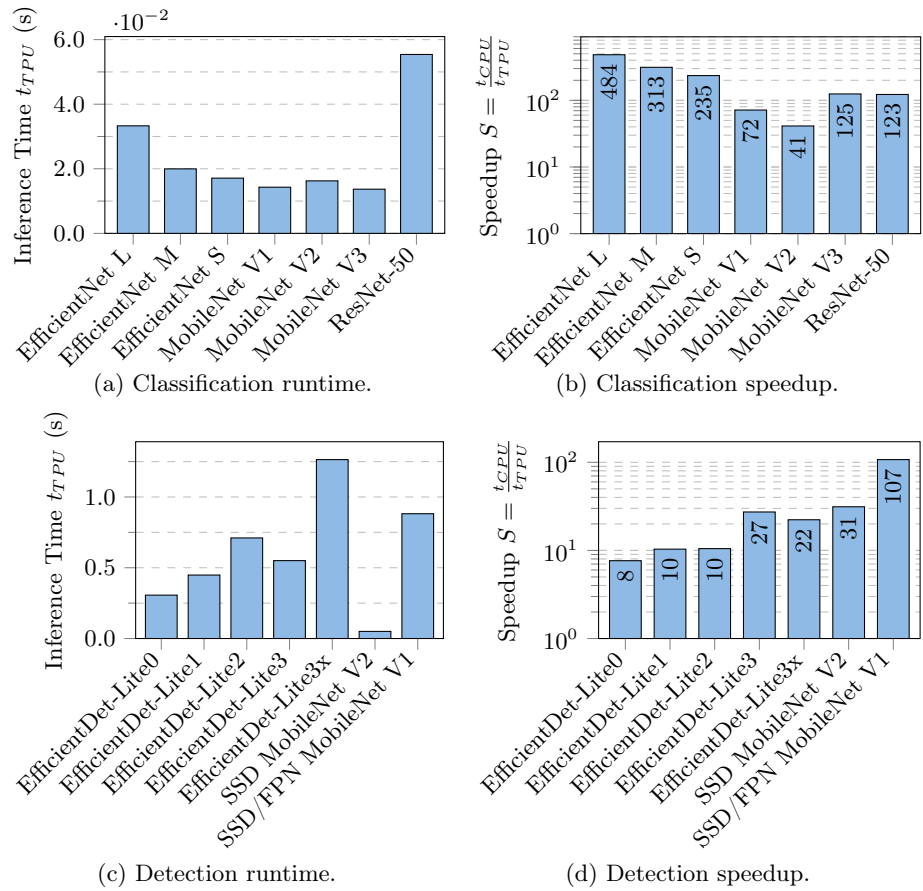
\begin{figure}[t!]

    \begin{subfigure}[b]{.5\textwidth}
        \pgfplotstableread[col sep=comma]{plot/data/classification.csv}\results
        \begin{tikzpicture}
    \pgfplotsset{every axis/.style={
                BarConfig,
                ybar,
                ymin=0,
                y tick label style={
                        /pgf/number format/.cd,
                        fixed,
                        fixed zerofill,
                        precision=1,
                        /tikz/.cd
                    },
                minor y tick num=1,
                yminorgrids=true,
                xticklabels from table={\results}{model},
                ylabel={Inference Time $t_{TPU}$ (\si{\second})},
                bar width=12pt,
            }}
    \begin{axis}
        \addplot[bar shift=+0pt, fill=rwth-50] table [x expr=\coordindex, y={avp64_edgetpu}] {\results};
    \end{axis}

\end{tikzpicture}
        \vspace{-1em}
        \caption{Classification runtime.}
        \label{fig:infdur:classtime}
    \end{subfigure}
    \begin{subfigure}[b]{.5\textwidth}
        \pgfplotstableread[col sep=comma]{plot/data/classification.csv}\results
        \begin{tikzpicture}
    \pgfplotsset{every axis/.style={
                BarConfig,
                ybar,
                ymin=1,
                ymode=log,
                yminorgrids=true,
                xticklabels from table={\results}{model},
                ylabel={Speedup $S = \frac{t_{CPU}}{t_{TPU}}$},
                bar width=12pt,
                transpose legend,
                legend pos={\legendposition},
                point meta=rawy,
                nodes near coords,
                every node near coord/.append style={rotate=90,anchor=east,/pgf/number format/.cd,fixed zerofill,precision=0},
            }}
    \begin{axis}
        \addplot[bar shift=+0pt, fill=rwth-50] table [x expr=\coordindex, y={avp64_speedup}] {\results};
    \end{axis}

\end{tikzpicture}
        \vspace{-2.2em}
        \caption{Classification speedup.}
        \label{fig:infdur:classspeed}
    \end{subfigure}
    \\
    \begin{subfigure}[b]{.465\textwidth}
        \pgfplotstableread[col sep=comma]{plot/data/detection.csv}\results
        \begin{tikzpicture}
    \pgfplotsset{every axis/.style={
                BarConfig,
                ybar,
                ymin=0,
                y tick label style={
                        /pgf/number format/.cd,
                        fixed,
                        fixed zerofill,
                        precision=1,
                        /tikz/.cd
                    },
                minor y tick num=1,
                yminorgrids=true,
                xticklabels from table={\results}{model},
                ylabel={Inference Time $t_{TPU}$ (\si{\second})},
                bar width=12pt,
            }}
    \begin{axis}
        \addplot[bar shift=+0pt, fill=rwth-50] table [x expr=\coordindex, y={avp64_edgetpu}] {\results};
    \end{axis}

\end{tikzpicture}
        \vspace{-2em}
        \caption{Detection runtime.}
        \label{fig:infdur:detecttime}
    \end{subfigure}
    \hspace{2.8mm}
    \begin{subfigure}[b]{.48\textwidth}
        \pgfplotstableread[col sep=comma]{plot/data/detection.csv}\results
        \begin{tikzpicture}
    \pgfplotsset{every axis/.style={
                BarConfig,
                ybar,
                ymin=1,
                ymode=log,
                yminorgrids=true,
                xticklabels from table={\results}{model},
                ylabel={Speedup $S = \frac{t_{CPU}}{t_{TPU}}$},
                bar width=12pt,
                transpose legend,
                legend pos={\legendposition},
                point meta=rawy,
                nodes near coords,
                every node near coord/.append style={rotate=90,anchor=east,/pgf/number format/.cd,fixed zerofill,precision=0},
            }}
    \begin{axis}
        \addplot[bar shift=+0pt, fill=rwth-50] table [x expr=\coordindex, y={avp64_speedup}] {\results};
    \end{axis}

\end{tikzpicture}
        \vspace{-2em}
        \caption{Detection speedup.}
        \label{fig:infdur:detectspeed}
    \end{subfigure}

    \caption{Inference results.}
    \label{fig:infdur}
    \vspace{-2em}
\end{figure}

\Cref{fig:infdur:classtime,fig:infdur:detecttime} show the wall-clock time needed to run the classification and detection \acp{nn} using the \ac{tpu} from the \ac{vp} with our hybrid approach.
The shown values are median values of \num{100} consecutive runs.
Due to the complexity of the tasks, the detection workloads take longer to execute than the classification ones.
\Cref{fig:infdur:classspeed,fig:infdur:detectspeed} present the speedups that result from the \ac{tpu} offloading compared to \ac{cpu}-only execution.
The results are obtained by dividing the wall-clock time needed for the \ac{cpu}-only execution, $t_{CPU}$, by the wall-clock time needed for inference with \ac{tpu} offloading, $t_{TPU}$.
In general, speedups between approximately \SI{10}{\texttimes} and \SI{480}{\texttimes} are reached.

\subsection{Communication Analysis}

A benefit of using \acp{vp} for \ac{sw} development compared to real \ac{hw} is the simplicity of tracing data collection.
Since the \ac{vp} offers insights into the models and their communication, complex debuggers or probes typically required for \ac{hw} analysis are no longer needed.
For SystemC-based simulation, external tools are available that make use of the standardized interfaces~\cite{bosbach_nistt_2022}.

\begin{figure}[!t]
    \centering
    \pgfplotstableread[col sep=comma]{plot/data/stats.csv}\results
    \begin{subfigure}[b]{\linewidth}
        \renewcommand{\legendposition}{north west}
        \begin{tikzpicture}
    \pgfplotsset{every axis/.style={
                BarConfig,
                ybar stacked,
                xmin=0,
                xmax=14,
                ymin=0,
                ymax=600,
                yminorgrids=true,
                xticklabels from table={\results}{model},
                ylabel={Data Accessed (Byte)},
                bar width=8pt,
                transpose legend,
                legend pos={\legendposition},
                enlarge x limits=0.1,
                legend style={legend columns=2},
            }}
    \begin{axis}[bar shift=-4pt]
        \addplot[fill=rwth-50] table [x expr=\coordindex, y={config_bytes_written_mean}] {\results}; \label{pgf:cfg_written}
        \addplot[fill=grun-50] table [x expr=\coordindex, y={bar_bytes_written_mean}] {\results}; \label{pgf:bar_written}
    \end{axis}
    \begin{axis}[hide axis, bar shift=+4pt]
        \addlegendimage{/pgfplots/refstyle=pgf:cfg_written}\addlegendentry{Config Write}
        \addlegendimage{/pgfplots/refstyle=pgf:bar_written}\addlegendentry{MMIO Write}

        \addplot[fill=rwth-50, postaction={pattern=north east lines, pattern color=black}] table [x expr=\coordindex, y={config_bytes_read_mean}] {\results};
        \addlegendentry{Config Read}
        \addplot[fill=grun-50, postaction={pattern=north east lines, pattern color=black}] table [x expr=\coordindex, y={bar_bytes_read_mean}] {\results};
        \addlegendentry{MMIO Read}
    \end{axis}

\end{tikzpicture}
        \vspace{-1em}
        \caption{Memory accesses.}
        \label{fig:stats:mem}
    \end{subfigure}
    \\
    \begin{subfigure}[b]{\linewidth}
        \renewcommand{\legendposition}{north west}
        \begin{tikzpicture}
    \pgfplotsset{every axis/.style={
                BarConfig,
                ybar stacked,
                yminorgrids=true,
                xticklabels from table={\results}{model},
                ylabel={Interrupt Counts},
                bar width=10pt,
                transpose legend,
                legend pos={\legendposition},
            }}
    \begin{axis}
        \addplot[fill=rwth-50] table [x expr=\coordindex, y={irq0_mean}] {\results};
        \addlegendentry{Instr. Queue}
        \addplot[fill=grun-50] table [x expr=\coordindex, y={irq4_mean}] {\results};
        \addlegendentry{SC-Host 0}
    \end{axis}

\end{tikzpicture}
        \vspace{-1em}
        \caption{Interrupt counts.}
        \label{fig:stats:irq}
    \end{subfigure}

    \caption{\acs{tpu} execution statistics.}
    \label{fig:stats}
    \vspace{-2em}
\end{figure}
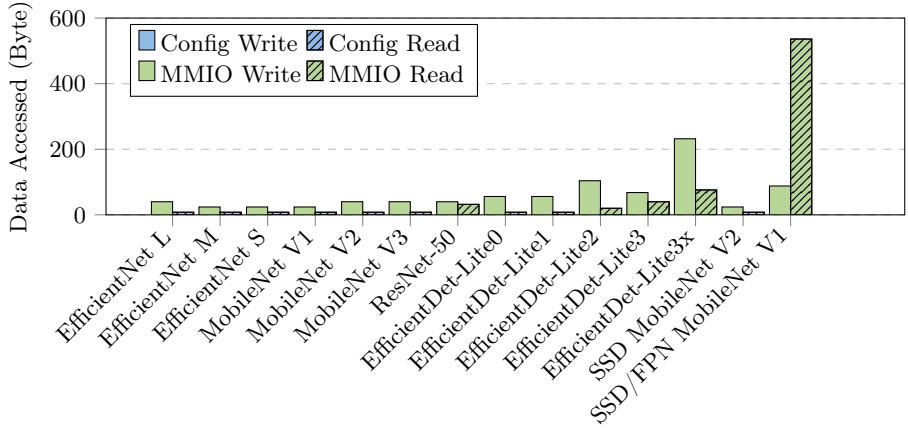
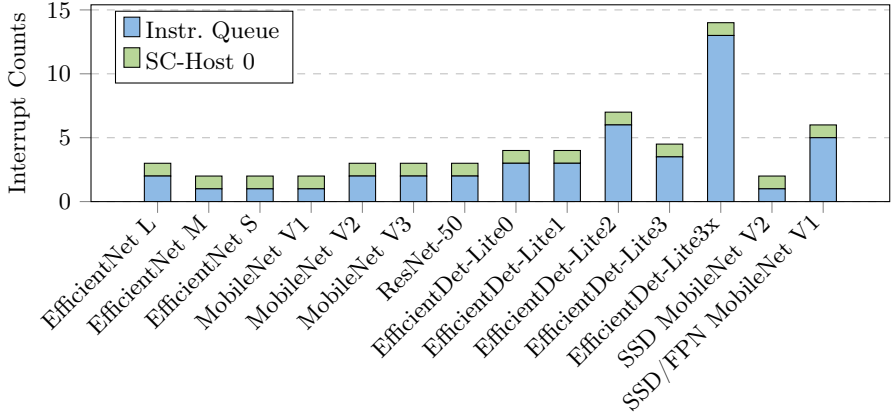

In this paper, we use the tracing capabilities of \ac{vcml} to log \ac{tlm} transactions from the \ac{vcpu} to the \ac{vpci} device, and \ac{msi}(-X) sent from the \ac{vpci} device.
In \cref{fig:stats:mem}, the number of bytes that are exchanged between the \ac{vcpu} and the \ac{vpci} device are shown.
The \ac{tpu} features a configuration and an \ac{mmio} region.
Since the configuration region is mainly used during driver initialization, it is not accessed during workload execution.
In general, accesses to the \ac{mmio} region are limited because it is only used for device configuration.
Data such as \ac{nn} models or inputs are accessed directly by the \ac{tpu} using its \ac{dma}.
These accesses cannot be traced because they are bypassing the \ac{vp} (see \cref{sec:implementation:memory}).
However, analyzing the \ac{vcpu}-to-\ac{vpci} communication provides valuable insights, particularly for debugging or profiling.

Further useful information can be obtained from the interrupt behavior.
The \ac{tpu} can send \num{13} different \ac{msi}-Xs.
For the executed workloads, only two of them, \ac{irq}0 and \ac{irq}4 (sc-host~0 \ac{irq}), are used.
\ac{irq}~0 (instruction queue) is signaled when the instruction queue has been executed by the \ac{tpu}.
\ac{irq}~4 (sc-host 0) is signaled once the execution has been completed.
\Cref{fig:stats:irq} shows the number of sent \acp{irq} for the different workloads.
For all workloads, the sc-host~0 \ac{irq} has been signaled once at the end of the execution.
Depending on the complexity of the workload, the instruction-queue \ac{irq} has been signaled multiple times.
Details about the \ac{irq} behavior, especially in combination with other tracing data like read/write accesses can help developers during \ac{sw} or driver development.
They provide valuable insights and can be used for automated testing.

\section{Conclusion and Future Work}
In this paper, we present an approach to integrate \ac{pci}(e) devices into a SystemC-\ac{tlm}-based \ac{vp}.
\ac{pci}(e) is today's standard for extension cards that can be connected to a general-purpose PC.
\acp{vp} allow \ac{sw} development for \acp{isa} other than the host machine.
This enables early \ac{sw} development even when the \ac{hw} is not available or still under design.
Our approach enables forwarding the access to a \ac{pci}(e) device that is connected to the host machine to a SystemC-based \ac{vp} running on the host.
This comes with multiple benefits such as increased performance, the superfluity of the creation of a virtual model, and simple regression testing.

The usage of SystemC \ac{tlm}-2.0 as the basis of our model and Linux's \ac{vfio} driver for accessing the \ac{pci}(e) devices guarantees portability to different \acp{vp} and \ac{pci}(e) devices.
Tracing data reveals the \ac{vcpu}-\ac{vpci} communication and the signaled \acp{irq}.
This information can help developers during driver or \ac{sw} development.

In a case study, we showed how our \ac{vpci} model can be used to embed Google Coral's Edge \ac{tpu} into an ARM-based \ac{vp}.
We presented how the model is integrated, accessed by the \ac{vcpu}, and how it signals interrupts and uses the host's \ac{iommu} to get direct access to the \ac{vp}'s \ac{ram}.
Performance results demonstrate that our integration can significantly accelerate a \ac{vp} by offloading \ac{ai} tasks to the \ac{tpu}.
This enhancement improves \ac{sw}-design productivity by accelerating simulation speeds by up to two orders of magnitude.

While this paper lays the foundation for integrating \ac{pci}(e) devices into SystemC-based \acp{vp}, there are several avenues for future research and development.
Due to the large number of available \ac{pci}(e) devices, our approach opens the doors for several scenarios.
In addition to \ac{ai} accelerators, other devices such as Ethernet or graphics cards can be used to replace virtual models.
This reduces the modeling effort and increases the simulation speed.
\newpage
\bibliographystyle{latex/splncs04}
\bibliography{library.bib}

\end{document}